\documentclass{aa}  
\pdfoutput=1 
\usepackage[colorlinks=true, allcolors=blue]{hyperref}
\hypersetup{colorlinks=true, linkcolor=black, urlcolor=red, citecolor=blue}
\usepackage{graphicx}
\usepackage{txfonts}
\usepackage{amsmath}
\usepackage{float}
\usepackage{siunitx}
\DeclareSIUnit{\Msun}{\ensuremath{M_\odot}}
\usepackage[dvipsnames]{xcolor}
\usepackage{physics}
\usepackage{pdflscape}
\usepackage{float}
\usepackage{afterpage}
\usepackage{csquotes}
\usepackage{nameref}
\usepackage{pdflscape}
\usepackage{subcaption}
\usepackage[table,xcdraw]{xcolor}
\usepackage{booktabs}
\usepackage{placeins}

\usepackage{gensymb}

\begin{document} 



\title{Alignment of the Milky Way and M31 with their cosmic environment}
\subtitle{New insights from constrained Local Group simulations}

\titlerunning{Alignment of the Milky Way and M31 with their cosmic environment}


\author{{Hanneke C. Woudenberg\inst{1}} 
          \and 
          {Amina Helmi\inst{1}}
          \and
          {Ewoud Wempe\inst{2}}
          \and
          {Anna Genina\inst{3, 4}}
          \and 
          {Simon D.M. White\inst{4}}
          \and 
          {Jens Jasche\inst{5}}
          \and 
          {Guilhem Lavaux\inst{6}}
          }

    \institute{Kapteyn Astronomical Institute, University of Groningen, Landleven 12, 9747 AD Groningen, the Netherlands
    \and
    LIRA, Observatoire de Paris, Université PSL, Sorbonne Université, Université Paris Cité, CY Cergy Paris Université, CNRS, Meudon, 92190, France
    \and 
    Institute for Astronomy, University of Edinburgh, Royal Observatory, Blackford Hill, Edinburgh EH9 3HJ, UK
    \and
    Max-Planck-Institut für Astrophysik, Karl-Schwarzschild-Straße 1, 85748 Garching, Germany
    \and
    The Oskar Klein Centre, Department of Physics, Stockholm University, Albanova University Center, SE 106 91 Stockholm, Sweden
    \and
    CNRS \& Sorbonne Université, UMR 7095, Institut d’Astrophysique de Paris, 98 bis boulevard Arago, F-75014 Paris, France}

   \date{Received xxxx; accepted yyyy}

\abstract
  {It is generally believed that the large-scale environment plays an important role in setting a galaxy's properties. Locally, the Milky Way (MW) and Andromeda (M31) reside in the Local Group, which itself is embedded in the Local Sheet.}
   {To study the influence of this sheet-like environment on the dark matter (DM) halo shapes, spins, and disks orientations of MW and M31 analogues, we use a new suite of constrained Local Group simulations {which reproduce the observed configuration of the two main halos and the Local sheet.}}
   {We have determined their shapes and alignments with respect to the  {Local Sheet analogues} in our simulations, as well as the effect of the infall  {and coalescence} of massive mergers.}
  {We find that the dark matter halos of our MW and M31 analogues have  {on average} slightly rounder shapes than reported in literature for similar mass galaxies in random environments. We also find  {preferential} alignment between the sheet normal and the halos’ minor axes, but not with the halos’ spins. The present-day disk angular momenta generally closely align with the halos' minor axes (median $15^{+15}_{-8}$  {degrees} at the virial radius)  {and also with the halos' spins}. We find that the direction of the disk angular momentum is  {often} set  {by one of the two highest mass ratio mergers }experienced in the last {8 -- 10} Gyr of evolution. While recent ($\leq 2$~Gyr) massive mergers can reorient the outer halo’s minor axis leading to twisted shapes, the disk angular momentum retains the imprint of the earlier accretion event.}
   {These results can explain the peculiar alignment of the Milky Way’s disk and DM halo shape, and their orientation with respect to the Local Sheet. The  {prolate-like morphology and orientation of the MW's outer halo} {can be }explained by the Magellanic Clouds (and perhaps Sagittarius) accreting from within the Local Sheet. {As their orbital planes are nearly perpendicular to the Galactic disk, this means that } the disk orientation  {must have been }set earlier, {possibly} by the Gaia-Enceladus-Sausage merger. {Interestingly, this merger's estimated infall direction, i.e. highly inclined with respect to the present-day Local Sheet, is broadly consistent with the Sheet's direction of maximum collapse.}}

\keywords{Local group — dark matter - Galaxy: kinematics and dynamics}

\maketitle  
\nolinenumbers

\section{Introduction} 
\label{sec:intro}
The distribution of galaxies in the Universe follows a  {web-like structure \citep{deLapparent1986, Bond1996, Gott2005}}. The cosmic web, as this is known, is composed of clusters, filaments, sheets and voids. Locally, the Local Group galaxies, i.e. our Milky Way (MW), Andromeda (M31), and their dwarf galaxy  {companions}, are embedded within the Local Sheet, a planar distribution of galaxies extending tens of Mpc bounded by the Local Void and the small Local Minivoid \citep{Peebles2001, Karachentsev2002, Tully2008, Peebles2010, McCall2014}. 

This sheet-like structure is expected to have shaped the orientation of the dark matter halos of the MW and M31 as well as of their disks, as  {satellites are} preferentially  {accreted} from within the sheet {at late times} \citep{Wempe2025b}. Statistical studies using cosmological N-body simulations seem to support a picture of preferential shape alignment, where halo minor axes  {of systems with stellar masses $M_\star \gtrsim 10^{10.5}$  \SI{}{\Msun} }tend to align perpendicular to sheets \citep{AragonCalvo2007, Codis2018}.
 {This shape alignment signal is found to be stronger than that from spin alignment. } {Indeed}, simulations show no clear preferred orientation of halo spins relative to sheet environments for MW-mass systems,  {as} the spin alignment transitions from preferentially parallel at lower masses ($M_{\rm halo} \lesssim 10^{11.5}$ \SI{}{\Msun}) to preferentially perpendicular at higher masses ($M_{\rm halo} \gtrsim 10^{12.5}$  \SI{}{\Msun})  \citep{Codis2014, Codis2018, Kraljic2020, AragonCalvo2020}. {A similar spin flip is seen for galaxy spins, which tend to align preferentially parallel and perpendicular to sheets  for $M_{\star} \lesssim 10^{9.5}$ \SI{}{\Msun}  and $M_{\star} \gtrsim 10^{10.5-11}$ \SI{}{\Msun}, respectively \citep{Codis2018, Kraljic2020}. This is supported by recent }observational work using the MANGA survey \citep{Moon2025}, which suggests  preferential alignment (at the $\sim 2 \rm \sigma$-level) of the galaxies' spins with the normal to the sheet for systems with  $M_\star \gtrsim 10^{10.5}$  \SI{}{\Msun},
 {although} earlier  {observational} work \citep[e.g.][]{Zhang2013,Tempel2013} yielded somewhat contradictory results.  {In the Local Group,} the spin  {vector} of the MW's disk is in the plane of the Local Sheet  {(within one degree\footnote{Assuming the north pole of the Local Sheet is located at Supergalactic $L =241.74 \degree$, $B= 82.05 \degree$ \citep{McCall2014}}}). M31's disk spin 
 {\citep[e.g.][]{Pawlowski2013b} is tilted by 145$\degree$ relative to north pole of the Local Sheet and} inclined by 77$\degree$ relative to the MW \citep{Corbelli2010}.
 
\begin{figure}[htb]
	\centering
	\includegraphics[width=\linewidth]{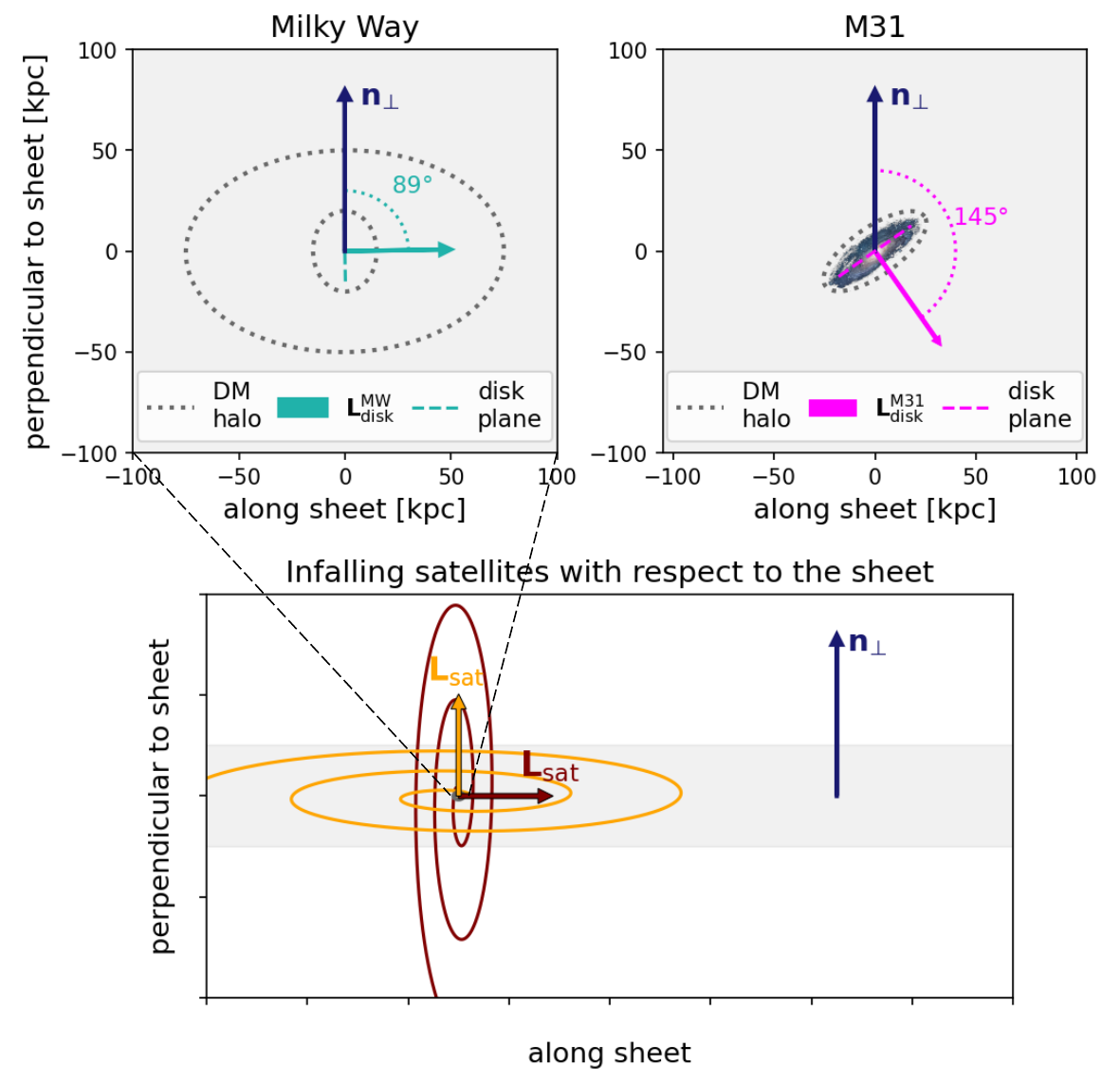} 
	\caption{\small \textbf{Top panels:} Schematic of the observed orientation of the MW
(left) and Andromeda (right) disk planes (dashed coloured lines), their
disk angular momenta (coloured arrows), and the estimated shape of their DM halos (dotted lines) relative to the Local Sheet.  {For simplicity, halo shapes are approximated as being axisymmetric (oblate or prolate), and we assumed that the plane of axisymmetry coincides with the disk plane. }The $y$-axis points in the direction perpendicular to the sheet (indicated by $\mathbf{n}_{\bot}$, dark blue arrow),  {while the $x$-axis lies in the sheet plane, oriented so that the disk appears edge-on}. M31 image credit: R. Gendler (2002). \textbf{Bottom panel}: Example of a satellite falling in from a direction 
within the sheet (orange) and perpendicular to the sheet (dark red), where the
arrows indicate the direction of their orbital angular momenta. }
	\label{fig:cartoon}
\end{figure}

Naturally, the dark matter (DM) halo shapes of the MW and M31 are less well constrained than their disk's spins. M31's DM halo has not been studied extensively, but might be oblate in the inner regions \citep[flattened towards the disk,][]{Banerjee2008}. On the other hand, the MW's DM halo shape has been the subject of many works, with its inner shape likely being oblate or nearly-spherical \citep[e.g.][]{Kupper2015, Malhan2019}, becoming more prolate and slightly triaxial around 15-20~kpc \citep{Bovy2016GD1Pal5, Posti2019, Dodd2022, Palau2022, Woudenberg2024, Nibauer2025}, and more strongly triaxial beyond, while still having its major axis broadly aligned with the normal to the disk  \citep{Lawmajewski2010, VeraCiro2011DMhalo, Dillamore2026, Li2026}. Furthermore, its minor axis appears to be within 30$\degree$ to 40$\degree$ of the orbital pole of the  Large Magellanic Cloud \citep[LMC][]{VasilievTango2021}, likely reflecting the effect of the infall of the LMC \citep[e.g.][]{Erkal2019, GaravitoCamargo2021}. Fig. \ref{fig:cartoon} shows a schematic representation of the MW's and M31's orientation with respect to the Local Sheet, revealing that the orientation  {of the MW's halo} could reflect the environmental influence of the Local Sheet.

Recently, \cite{Wempe2024} presented a  {posterior sample} of $\Lambda$CDM Local Group simulations that were constrained to reproduce the MW's and M31's masses, relative velocities, and positions, as well as the quiet Hubble flow observed for galaxies out to 4~Mpc. 
 {As a result of the dynamical constraints imposed on 1 to 5~Mpc scales, these simulations consistently produce a sheet-like structure surrounding the Local Group with an orientation close to that of the actual Local Sheet.}
In these simulations, the MWs  {tend to} assemble their mass earlier than the M31s, evolving more quiescently at late times \citep{Wempe2025a}, in rough agreement with the observations \citep{Helmi2018, Belokurov2018,Bell2018}. {This suggests that it is important to take into account the local environment when attempting to understand properties of the Milky Way.}

In this work, we investigate the orientation, spins and shapes of the MW and M31 analogues in a suite of dark matter only (DMO) and hydrodynamical simulations using the  {initial conditions inferred in} \cite{Wempe2024}. We aim to understand how the Milky Way obtained its current disk orientation and alignment with its DM halo.
Our paper is structured as follows.  Sect.~\ref{sec:datamethod} presents the simulations used and in Sect.~\ref{sec:shapes} we determine the DM halo shapes, their angular momenta as well as that of the disks, and the orientation of the sheet plane. In Sect. \ref{sec:results}  we study the alignments with respect to the environment as well as their drivers. In Sect.~\ref{sec:disc}  we discuss our findings and  {their} implications for the MW, and  {we} present our conclusions in Sect.~\ref{sec:conclusion}.

\section{Description of the simulations} \label{sec:datamethod}

\cite{Wempe2024} extended the Hamiltonian Monte Carlo-based {\sc Borg} methodology \citep[e.g.][]{Jasche2013, Lavaux2016, Jasche2019} to the scales relevant to the 
Local Group by inferring initial conditions in the
context of $\Lambda$CDM, which when evolved to the present day satisfy a priori specified constraints on the masses of the MW and M31, their relative motions,  {and the observed recession velocities of surrounding isolated galaxies out to $\sim 4$~Mpc. The latter constraint ensures that the simulated Local Groups are typically embedded in a Local Sheet whose orientation closely matches that observed \citep{Wempe2026}.}  {The initial condition inference used quick, low-resolution cosmological DMO simulations} of a 10~Mpc zoom region centred on the Local Group embedded in a 40~Mpc low-resolution box with a particle mass of  \SI{1.51 e8}{\Msun} and force mesh voxel size of 78.125~kpc in the high resolution region. A number of the inferred initial conditions have been re-simulated using {\sc Gadget-4} \citep{Springel2021} at  {various higher} resolutions \citep[DMO LR and DMO HR described in][respectively]{Wempe2025a,Wempe2025b}, and a subset of those have also been simulated including baryonic physics \citep[hydro LR and hydro HR, described in][]{Geninainprep} using the {\sc Arepo} code \citep{Springel2010, Weinberger2020} with the {\sc Auriga} galaxy formation model \citep{Grand2017}. Table~\ref{Table:props_sims} provides an overview of the simulation properties. 

Subhalo catalogues and merger trees for the DMO and hydro simulations were constructed from the high resolution particles using SUBFIND-HBT  {integrated within {\sc Gadget-4}} \citep{Han2018, Springel2021}. We use here the infall catalogues of \cite{Wempe2025b}, which track the properties of (sub)halos  for each branch of the merger tree (i.e. peak mass, first FoF infall, first $R_{\rm vir}$ crossing\footnote{In this paper, we define $R_{\rm vir}$ as the radius of a sphere with an average density of $200 \rho_c$, where $\rho_c$ is the critical density of the Universe. However, note that for the subhalo catalogues a slightly different definition is used, such that $R_{\rm vir}^\text{approx} = \sqrt[3]{M_{\rm halo}/(\frac{4}{3} \pi 200 \rho_{\rm crit})}$ {, where $M_{\rm halo}$ is the mass of the main subhalo. This approximation agrees well with the spherical-overdensity $R_\text{vir}$ measured from the particle distribution, with $R_\text{vir}^\text{approx}/R_\text{vir}=0.99\pm0.04$ over all simulations and snapshots.}}, and final snapshot before destruction), and we compute similar infall catalogues for the hydrodynamical runs.

\begin{table}[t!]
\centering
\caption{ \small Overview of the properties of the simulations studied in this work. We quote the mass and softening ($\epsilon$) of the high resolution DM particles part of the zoom region. For the hydro-simulations we quote the average stellar particle mass at the present day and the Plummer-equivalent softening.  {We tested that using a larger softening for the LR runs did not  {change} our results. } Three hydrodynamical runs were additionally run using the LR mass resolution and HR softening length (\texttt{982\_1629}, \texttt{983\_1809}, \texttt{986\_1219}), making  {them the lowest resolution runs}.  }
\label{Table:props_sims}
\begin{tabular}{llllll}
\toprule
  & $N_{\rm sim}$ &  {$m_{\rm DM}$ [ \SI{}{\Msun}]} & {$\epsilon_{\rm DM}$ [kpc]} &  $m_{\star}$  [ \SI{}{\Msun}]   \\     
\midrule
DMO LR     & 240          &  $1.9 \cdot 10^7$   &  2.22         & -       \\ 
DMO HR     & 86           &  $2.4 \cdot 10^6$   & 0.56          & -   \\ 
Hydro LR   & 9            &  $1.6 \cdot 10^7$   & 0.37          & $2.1 \cdot 10^6$  \\ 
Hydro HR   & 4            &  $2.0 \cdot 10^6$   & 0.74          & $2.5 \cdot 10^5$\\ 
\bottomrule
\end{tabular}
\end{table}

\section{Halo shapes, angular momentum and sheet  {orientation}}
\label{sec:shapes}

 {Here, we characterise the shapes and angular momenta of the MW and M31 DM halos and, for the hydrodynamical runs, we examine their alignment with the stellar disk. Next, we determine the orientation of the simulated Local Sheet analogues and examine their evolution over time.}

\subsection{DM halo shapes and angular momentum}

We determine the shape of a DM halo  {(using high resolution DM particles that are part of the main halo, excluding satellites)} through an iterative approach, following \cite{Allgood2006} and \cite{VeraCiro2011DMhalo}, using the reduced inertia tensor
\begin{equation}
    I_{ij} = \sum_{{\tilde{x}_k} \in V} \frac{{\tilde{x}}_k^{(i)} {\tilde{x}}_k^{(j)} }{d_k^2}, 
    \label{eq:I}
\end{equation}
where $d_k$ is the distance measured from the centre of the halo to the $k$th particle, defined as $d_k^2 = {\tilde{x}}_k^2 + {\tilde{y}}_k^2/p^2 + {\tilde{z}}^2_k/q^2$, and $V$ is the volume containing a set of particles' positions. $\tilde{x}_k$, $\tilde{y}_k$, and $\tilde{z}_k$ are aligned with the major ($\mathbf{a}$), intermediate ($\mathbf{b}$), and minor ($\mathbf{c}$) axes of the ellipsoid. Hence, $p = |\mathbf{b}|/|\mathbf{a}|$ is the ratio between the intermediate and major axis  {lengths}, and $q = |\mathbf{c}|/|\mathbf{a}|$ is the ratio between the minor and major axis  {lengths}. We keep the length of the largest axis, $\tilde{x}$, fixed, and reshape the ellipsoid $V$ containing the particles iteratively. We define convergence in $p$ and $q$ when a relative change smaller than $10^{-5}$ is reached. We make sure that at least 1000 particles are present within the smallest radius considered  {by requiring a minimum radius of at} least five times the softening length. Throughout this paper, unit vectors will be denoted with the $\: \hat{} \: $ symbol, and we will denote the direction of the halo minor axis direction  {(at some fixed radius)} as $\mathbf{\hat{c}}$ and its direction at the virial radius as $\mathbf{\hat{c}}_{\rm R_{\rm vir}}$.

\begin{figure}
	\centering
	\includegraphics[width=0.95\linewidth]{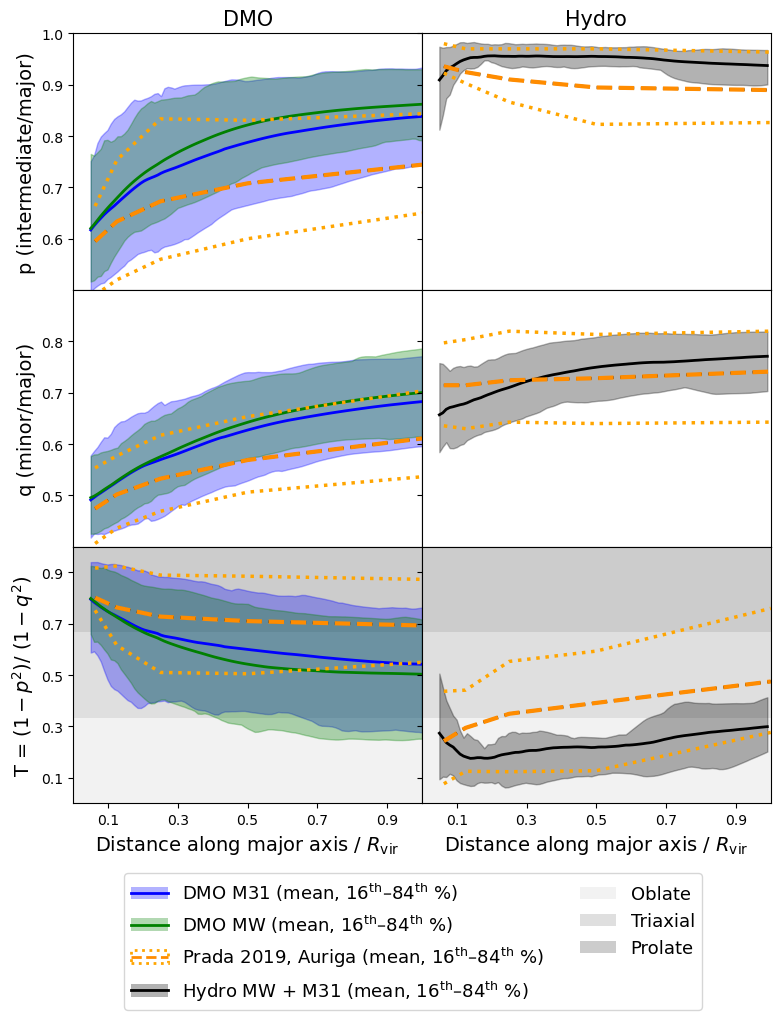} 
	\caption{\small {\bf Left column}: Shapes of the MW (green) and M31 (blue) halos in the HR DMO simulations (86 in total) as a function of distance along the major axis normalized by the halos' virial radii.  {The bottom row shows triaxiality, where $0 \leq T \leq 1/3$ corresponds to oblate shapes, $1/3 \leq T \leq 2/3$ corresponds  to triaxial shapes, and $2/3 \leq T \leq 1$ corresponds to prolate shapes.} The solid lines show the mean of the sample, the light shaded areas correspond to the 16th to 84th percentile range.  {The orange dashed and dotted lines indicate the mean shapes and 16th to 84th percentile range, respectively, found by \citet[their Table 1 and 2]{Prada2019}.} {\bf Right column:} Shapes of the DM halos of both the MWs and M31s in the hydrodynamical simulations (in black). }
	\label{fig:shapes_MW_M31_Rvir}
\end{figure}

Applying this methodology on the HR DMO runs, we find that the halos are triaxial  (see  Appendix~\ref{sec:appendix:qsprofiles} for detailed figures) and that there is no statistically significant difference between the MWs and M31s shapes, although the MW halos appear to be slightly more axisymmetric. Overall we find the halo shapes of our MWs and M31s to be  {triaxial and less prolate} than reported in the literature for similar mass systems, as shown in the left column of Fig.~\ref{fig:shapes_MW_M31_Rvir}. Here, we overplot the mean shapes (obtained using the same method)  {from \cite{Prada2019}, who used the set of 30 MW-like halos of the Auriga Level-4 simulations. These halos were selected to be relatively isolated at redshift zero, but otherwise inhabit random environments \citep[for more details, see][]{Grand2017}.} 
Our mean curves for shape parameters $q$ and $p$ lie systematically  {above those of \cite{Prada2019}, with the offset increasing at larger radii}. {To assess if this difference is  statistically significant, we compare the distributions of $q$ and $p$ at $R = R_{\rm vir}$ 
using a bootstrapped KS test (\texttt{scipy.stats.ks\_2samp}) with $10^4$ iterations, where we draw a subsample of $N=30$ without replacement from our data}. {When comparing to the \cite{Prada2019} sample, we find KS $p$-values $< 3\times10^{-3}$ for  the distributions of $p$, and 
$< 9\times10^{-5}$ for $q$. At smaller radii,
no significant difference is detected.}
In the hydrodynamical runs (right panels of Fig.~\ref{fig:shapes_MW_M31_Rvir}), the shapes of the DM halos become oblate (close to being axisymmetric with $p$ approaching unity) in the inner regions, while in the outer regions they tend to become triaxial, but  {again} to a lesser degree than found in previous works \citep[e.g.][and references therein]{Chua2019, Prada2019}.  {However, this difference is not statistically significant}.

We also computed the DM halos' total angular momentum {(using high resolution  DM particles that are part of the main halo, excluding satellites)} with respect to the halo's central position and velocity within volume $V$, defined as 
\begin{equation}
    \mathbf{J}_{\rm tot} = \sum_{i \in V} \mathbf{r}_i \times \mathbf{p}_i .
    \label{eq:J}
\end{equation}
We find good convergence between the LR DMO runs, HR DMO runs and hydrodynamical runs for the DM halos' shape and for the angular momentum $\mathbf{J_{\rm tot}}$ as a function of radius at the present-day (see Appendix \ref{sec:appendix:convergence}). {We also find that the DM halos' minor axis directions $\mathbf{\hat{c}}$ and their angular momentum $\mathbf{J_{\rm tot}}$ are clearly correlated, but the scatter is large}. 

To track the direction of $\mathbf{L}_{\rm disk}$ in the hydrodynamical runs over time, we compute, for each snapshot, the mean angular momentum using stars younger than 3 Gyr and located within 10 kpc of their halo centres. We confirmed that the disk orientation converges well between the LR and HR hydro simulations  {(within $14.5^{+15.6}_{-10.1}$ degrees at the present day)}. We find good alignment between $\mathbf{L}_{\rm disk}$ with both  {the minor axis direction} $\mathbf{\hat{c}}$ and  {total angular momentum} $ \mathbf{J}_{\rm tot}$. At the virial radius, $\mathbf{L}_{\rm disk}$ lies within $15^{+ 15}_{- 8}$  {degrees} of $\mathbf{\hat{c}}_{\rm R_{\rm vir}}$ and within $25^{+20}_{-16}$  {degrees} of $\mathbf{J}_{\rm tot}^{\rm R_{\rm vir}}$ ( {having a} larger scatter, see also Fig. \ref{fig:disk_ang_mom_minor_axis},  {but note that the possible misalignment also spans a twice as large range}). In the inner regions ($r \lesssim 50$ kpc), these alignments are generally stronger. Systems undergoing a massive merger (with mass $\geq$  \SI{e11}{\Msun} at the present-day) can  {show} a misalignment of many tens of degrees,  especially at large radii. This is the case for some of our simulations (M31 720, MW 910 or MW 1540). The average degree of alignment found is consistent with, but slighter tighter than, reported in the literature \citep[e.g.][]{Deason2011, Teklu2015, Valenzuela2024}. 

\subsection{Orientation of the sheet}
\label{sec:sheet-orient}
Observationally, the Local Sheet is defined using the distribution of galaxies  \citep[e.g.][]{Tully2008, McCall2014}. To determine the orientation of the sheet in  each of our simulations, we take the positions of the most bound particles of all subhalos with masses larger than  \SI{e8}{\Msun} {within a sphere} of radius 5~Mpc centred on the MW and assume that locally the sheet can be approximated by a plane. We then fit a plane (defined by a point and a normal) to the distribution of halo positions using \texttt{skspatial} which performs a singular value decomposition. 

\begin{figure}
	\centering
	\includegraphics[width=\linewidth]{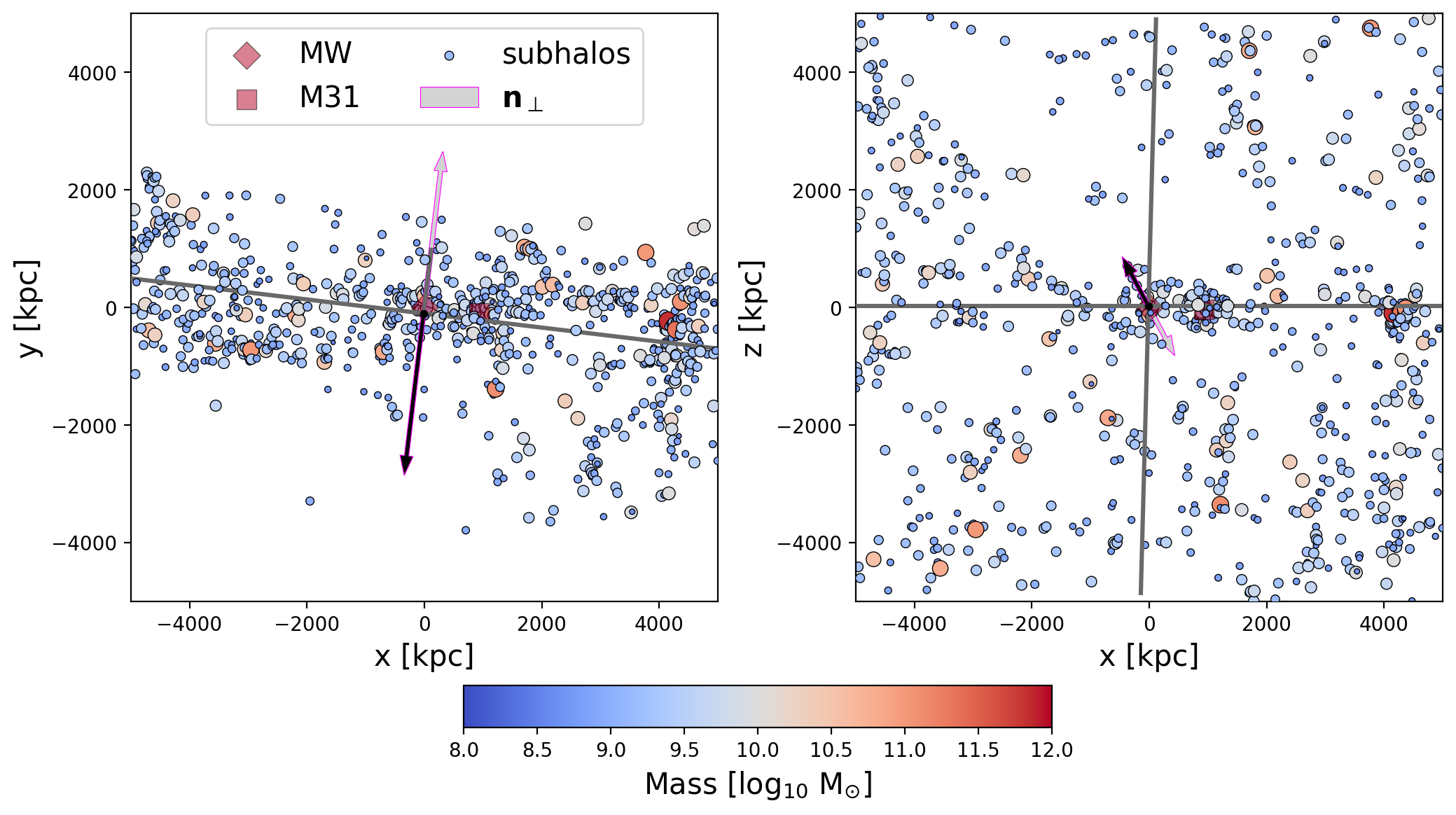} 
	\caption{\small Distribution of  {DM subhalos} with masses $ > $  \SI{e8}{\Msun} in LR DMO run \texttt{98.10\_mcmc\_1019} within a cube of 10~Mpc on a side centered on the MW in the simulation coordinates. The DM halos are color-coded and sized by their mass. The orientation of the sheet is obtained by fitting a plane to the halos' distribution. The grey lines indicate the orientation of the plane, shown in  $(x,y)$ (nearly edge-on, left panel) and $(x,z)$ (nearly face-on, right panel) projections, respectively. The light magenta arrow indicates the direction of the normal to the sheet, $\mathbf{n}_{\bot}$, the dark magenta arrow $- \mathbf{n}_{\bot}$. 
    }
	\label{fig:example_fit_sheet}
\end{figure}

We find that the sheet is well captured by this method, see Fig. \ref{fig:example_fit_sheet} for an example. The direction perpendicular to the sheet, $\mathbf{n}_{\bot}$, is typically closely aligned with the simulations' $y$-axis (median alignment \SI{\sim 10}{\degree}). Since the normal is defined up to a sign, we choose the convention that $\mathbf{n}_{\bot}$ points  {in the direction of} the positive $y$-axis, making it roughly align with what observationally would be the North Supergalactic Pole.  {Note that, as discussed by \cite{Wempe2026}, this is a highly non-trivial result since the {\it positions} of galaxies beyond the Local Group were not used in creating the initial conditions for our simulations, only the 
Hubble flow which they delineate.} 

We tested the robustness of our sheet-fitting method as follows.  {We find that including the low-resolution particles outside the zoom region in the fit (each included individually, as their mass of $9.65 \cdot 10^9 M_{\odot}$ exceeds our subhalo mass threshold) }does not have a significant impact, as their distribution is roughly isotropic around the sheet and hence introduces a median difference of only \SI{\sim 5}{\degree},  {and this reduces to \SI{\sim 4}{\degree} when the mass-weighted reduced inertia tensor (one iteration) is used}). 
Furthermore, the derived $\mathbf{n}_{\bot}$ remains similar when fitted to spheres of radius 3.5 -- 8 Mpc, and the results do not vary by more than $1 \degree$ when changing the resolution or  including the effect of baryons.  {Moreover, we have tested that only taking the most bound particle of a FoF group gives a similar orientation within 0.4 degrees.}  {Finally, we  tested the effect of restricting the fit to halos with at least one star particle in the hydrodynamical runs, and find a difference of median 3 degrees (2 degrees for the HR~runs).}

We also tracked the evolution of the sheet over time for all LR DMO runs. To study its shape, we fit the particle distribution within a  {sphere of radius} 5~cMpc  {using the reduced inertia tensor (one iteration) including both the low and high resolution particles, and we fit the plane of the sheet using the method presented earlier.} We find that the sheet is largely in place from \SI{\sim 3}{}~-~\SI{4}{Gyr} onwards, with a thickness of approximately 0.7 cMpc (corresponding to $\pm 1\sigma$ from the plane). 
At early times (\SI{\sim 3}{Gyr}), the sheet normal is misaligned by a median of  {$7.7^{+4.3}_{-4.0}$ degrees} relative to the present-day $\mathbf{n}_{\bot}$, and this misalignment gradually diminishes in time.

\section{Alignments of halos and disks with respect to the environment and the role of massive mergers} \label{sec:results}

 {In this section we investigate the alignment of the DM halo minor axes and disk angular momenta with respect to the sheet environment, and examine how the infall of massive mergers influences these orientations.}

\subsection{Alignments with respect to the Local Sheet} \label{sec:results:shape_sheet}

\begin{figure}[t]
	\centering
	\includegraphics[width=\linewidth]{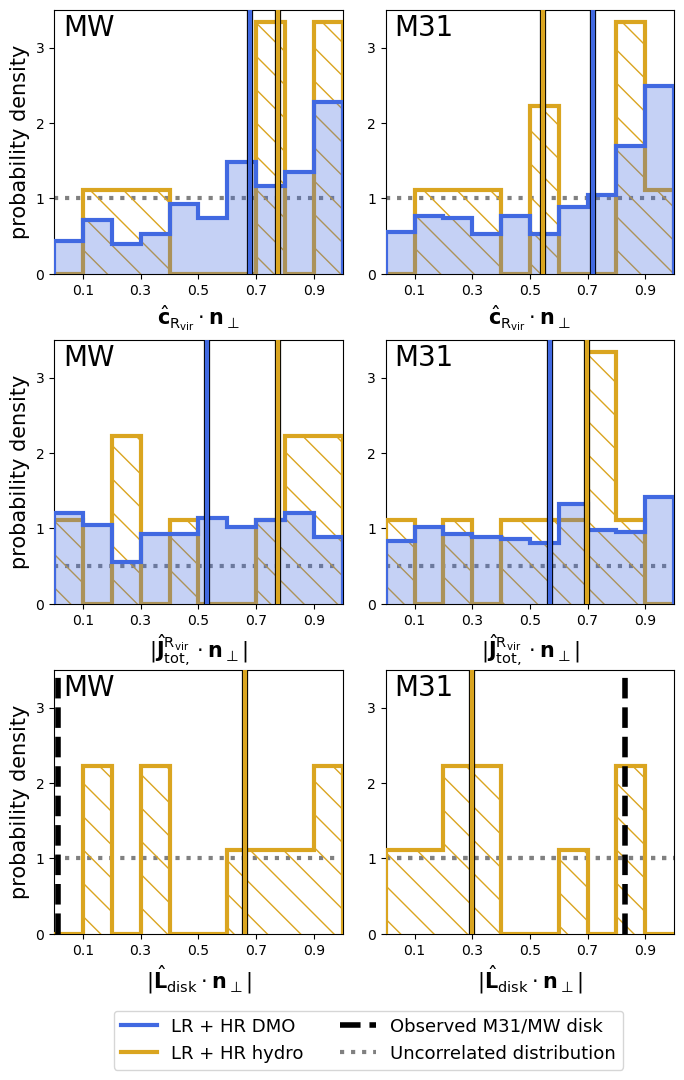} 
	\caption{\small \textbf{Top panel:}  {Probability} density distributions of the dot product between the direction of the minor axis ($\mathbf{\hat{c}}_{\rm R_{\rm vir}}$) and $\mathbf{n}_{\bot}$, for the LR DMO and HR DMO runs (green) and LR and HR hydro runs (yellow). The median of each distribution is shown as a solid line. An  {uncorrelated} distribution would appear uniform in this space,  {as illustrated by the grey dotted line}. The left panel shows the distributions for all MWs, the right panel shows the distributions for all M31s.  {If we would extrapolate the shape measurements for the MW and M31 to the virial radius, this would give $\mathbf{\hat{c}}_{\rm R_{\rm vir}}^{\rm MW} \cdot \mathbf{n}_{\bot} \sim 1$ and $\mathbf{\hat{c}}_{\rm R_{\rm vir}}^{\rm M31} \cdot \mathbf{n}_{\bot} \sim 0.8$.} \textbf{Middle panel:} Same as the top row, but for the density distributions of the absolute value of the dot product between the DM halo angular momentum at the virial radius ($\mathbf{\hat{J}}_{\rm tot, \: R_{\rm vir}}$) and $\mathbf{n}_{\bot}$. \textbf{Bottom panel:} Same as the top row, but for the density distributions of the absolute value of the dot product between the direction disk angular momentum $\mathbf{\hat{L}}_{\rm disk}$ and $\mathbf{n}_{\bot}$ for the LR and HR hydro runs. The observed angles between $| \mathbf{\hat{L}}_{\rm disk} \cdot \mathbf{n}_{\bot}|$ for the MW and M31 are indicated by black dashed lines.}
	\label{fig:minor_NGP_angle}
\end{figure}

Here, we examine whether the local environment induces a preferred alignment of the DM halo orientations, their angular momenta, and those of the disks with respect to the Local Sheet. We compute the angle between  {the minor axis direction at the virial radius} $\mathbf{\hat{c}}_{\rm R_{\rm vir}}$  and $\mathbf{n}_{\bot}$ for the LR DMO, HR DMO, LR hydro and HR hydro runs\footnote{We find no strong preferred alignment between $\mathbf{\hat{c}}_{\rm R_{\rm vir}}$ and the direction toward M31 or the MW, with a median value of $\mathbf{\hat{c}}_{\rm R_{\rm vir}} \cdot \mathbf{\hat{r}}_{\rm MW/M31} \sim 0.4$ in both DMO and hydrodynamical runs. This is comparable to $\mathbf{\hat{L}}_{\rm disk, MW} \cdot \mathbf{\hat{r}}_{\rm M31} \sim 0.4$ and $\mathbf{\hat{L}}_{\rm disk, M31} \cdot \mathbf{\hat{r}}_{\rm MW} \sim 0.2$.}. The top row of Fig. \ref{fig:minor_NGP_angle} shows that there is a preference for $\mathbf{\hat{c}}_{\rm R_{\rm vir}}$ to be aligned with $\mathbf{n}_{\bot}$, with a median value of $\mathbf{\hat{c}}_{\rm R_{\rm vir}} \cdot \mathbf{n}_{\bot} \sim 0.7$ for the DMO runs\footnote{A KS test (\texttt{scipy.stats.ks\_2samp})   {finds no statistically significant difference} between the LR and HR DMO distributions of $\mathbf{\hat{c}}_{\rm R_{\rm vir}} \cdot \mathbf{n}_{\bot}$, with KS statistics of 8.2\% ($p$ = 0.76) and 12.3\% ($p$ = 0.26) for the M31 and MW analogues, respectively. Both LR and HR $\mathbf{\hat{c}}_{\rm R_{\rm vir}} \cdot \mathbf{n}_{\bot}$ distributions are statistically distinct from a uniform distribution, with KS-statistics $\geq 18\%$ and $p \leq 0.03$.}, with the hydrodynamical runs following a similar trend.  {This is consistent with the preferential alignment found in e.g. \cite{AragonCalvo2007, Codis2018}}. 

 {The middle row of Fig. \ref{fig:minor_NGP_angle} shows }that the  {distribution of total angular momentum at the virial radius with respect to the sheet}, $|\mathbf{\hat{J}}_{\rm tot}^{\rm R_{\rm vir}} \cdot \mathbf{n}_{\bot}|$,  is not statistically different from a uniform distribution\footnote{{A KS test finds no statistically significant difference between the LR and HR DMO distributions of $|\mathbf{\hat{J}}_{\rm tot}^{\rm R_{\rm vir}} \cdot \mathbf{n}_{\bot}|$. Both the M31 LR, M31 HR, MW LR, and MW HR distributions are not found to be statistically different from a uniform distribution (KS-statistic $\leq$ 10\% and $p \geq 0.20$)}},
 with median values of $|\mathbf{\hat{J}}_{\rm tot}^{\rm R_{\rm vir}} \cdot \mathbf{n}_{\bot}| \sim 0.5$ for the MW analogues and $\sim 0.6$ for the M31 analogues (compared to $\sim 0.8$ and $\sim 0.7$, respectively, for the hydrodynamical runs). Our findings are in agreement with e.g. \cite{Codis2018, Kraljic2020, AragonCalvo2020}. 

The bottom panel of Fig. \ref{fig:minor_NGP_angle} shows the distribution for the cosine of the angle between $\mathbf{\hat{L}}_{\rm disk}$ and $\mathbf{n}_{\bot}$ for the hydrodynamical runs.  {Given the strong alignment between $\mathbf{\hat{L}}_{\rm disk}$ and minor axis direction $\mathbf{\hat{c}}_{\rm R_{\rm vir}}$ and the similarly strong alignment between $\mathbf{\hat{L}}_{\rm disk}$ and the total angular momentum (see also Appendix \ref{sec:appendix:convergence:minoraxis_Ldisk}), one would expect that the distribution of $|\mathbf{\hat{L}}_{\rm disk} \cdot\mathbf{n}_{\bot}|$ resembles those shown in the top two rows of the Figure.} Indeed, the median alignment for the MWs  is $|\mathbf{\hat{L}}_{\rm disk} \cdot \mathbf{n}_{\bot}| =0.7$, while that for the M31s is 0.3.
 {These values are compatible with the shape alignment and angular momentum alignment results shown in the top panels given the small number statistics. } 

For both the MW and M31 there are simulations in our suite with similar disk orientations as observed, as can be seen from Fig. \ref{fig:minor_NGP_angle}.  {However, the observed orientation of the MW's disk with respect to the Local Sheet, being $\mathbf{\hat{L}}_{\rm disk, MW} \cdot \mathbf{n}_{\bot}~\sim~0.0$, appears atypical compared to the average. The two MW analogues that share this alignment 
(MW 720 and MW 1330) however also show, like most systems in our sample, strong alignment between $\mathbf{\hat{L}}_{\rm disk}$ and  $\mathbf{\hat{c}}_{\rm R_{\rm vir}}$ (see Fig. \ref{sec:appendix:convergence:minoraxis_Ldisk}), in contrast to the observational evidence on the Milky Way. 
This implies that our hydrodynamical sample does not contain a fully realistic MW analogue.}

\subsection{Role of massive mergers in setting the DM halo and disk orientation}

While the observed orientation of the MW's minor axis at larger radii is consistent with the simulations, its disk orientation  {and twisted DM halo remain peculiar}. To better understand what could cause such an alignment  {between} the disk and its host DM halo, we examine the effect of massive mergers. We first investigate the link between the present-day alignment of $\mathbf{\hat{c}}_{\rm R_{\rm vir}}$ and the angular momentum direction of the three mergers {with the highest mass ratio} in both the DMO and hydrodynamical runs.  {We focus on the three highest mass ratio events as those will likely have had the largest impact on their host galaxy \citep[e.g.][]{Dodge2023, Bell2026}.} These mergers were selected by requiring they are  destroyed {after} the first 2~Gyr of evolution (i.e. less than 11.7 Gyr ago)\footnote{{We define the destruction time of a halo as the moment it disappears from the subhalo catalogue, and require destruction times smaller than 11.7~Gyr ago as we found that the effect of earlier mergers is washed out by later events.}}, a  first $R_{\rm vir}$ crossing more than 1~Gyr ago, and a mass ratio $\mu = M_{\rm sat}(t_{\rm peak})/ M_{\rm host}(t_{\rm peak}) > 0.08$, where $M_{\rm sat}$ and $ M_{\rm host}$ are the mass of the subhalo and host galaxy, respectively, both evaluated at $t_{\rm peak}$, the time the subhalo reaches its peak mass (i.e. just before infall). We recorded the angular momentum direction of these subhalos, $\mathbf{\hat{L}}_{\rm sat, infall}$,  at their first $R_{\rm vir}$ crossing (as then it likely has the largest impact on the host galaxy, since $\mathbf{L}$ scales with mass, radius and velocity), and computed its angle with respect to $\mathbf{n}_{\bot}$ to distinguish infall from within or outside the sheet. As illustrated in the bottom panel of Fig. \ref{fig:cartoon}, an angle close to $0 \degree$ ($\cos \sim 1$) corresponds to infall from within the sheet (orange orbit), while an angle close to $90 \degree$ ($\cos \sim 0$) to infall perpendicular to the sheet (dark red orbit). A random distribution would have an average infall angle of  {$60\degree$} ($\cos = 0.5$).

\begin{figure}[t!]
	\centering
	\includegraphics[width=\linewidth]{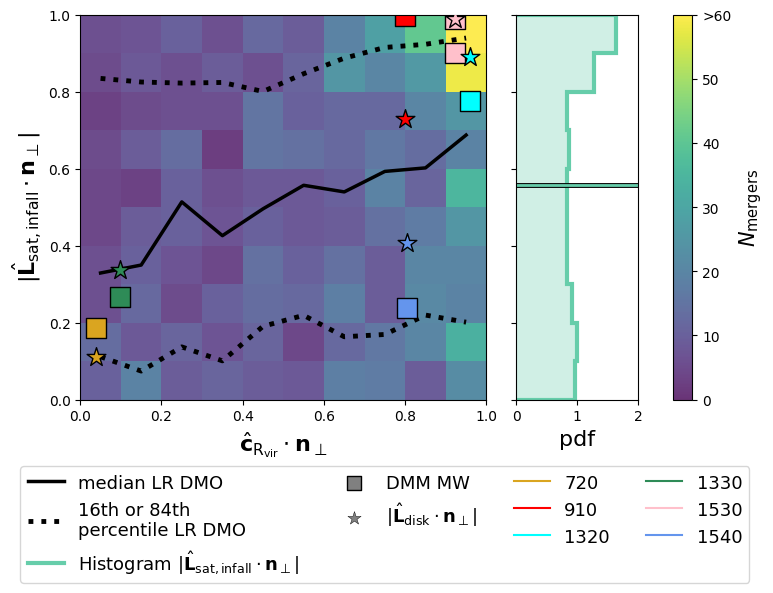} 
	\caption{\small \small \textbf{Left panel:} 2D histogram of $\mathbf{\hat{c}}_{\rm R_{\rm vir}} \cdot \mathbf{n}_{\bot}$ versus the cosine of the angle between the angular momentum direction at the moment of infall $\mathbf{\hat{L}}_{\rm sat, infall}$ and $\mathbf{n}_{\bot}$ determined from the three most massive mergers with $\mu \geq 0.08$ for all MWs and M31s in the LR DMO runs. When $|\mathbf{\hat{L}}_{\rm sat, infall} \cdot \mathbf{n}_{\bot}|$=1, the satellite fell in from within the plane of the sheet, while a value of 0 corresponds to infall perpendicular to it. The median is indicated as a solid black line, the 16$^{\rm th}$ and 84$^{\rm th}$ percentiles as dotted black lines. With coloured symbols we have overplotted for the MWs in our hydrodynamical runs, where the squares indicate $|\mathbf{\hat{L}}_{\rm sat, infall} \cdot \mathbf{n_{\bot}}|$ for the DDMs, while the  {$y$-axis coordinate of the star symbols is} $|\mathbf{L}_{\rm disk} \cdot \mathbf{n}_{\bot}|$, i.e. the orientation of the disk with respect to $\mathbf{n}_{\bot}$ at the present day.  {In the case of MW 1530, the disk orientation is determined by the averaged influence of two mergers with very similar angular momentum occurring within \SI{250}{Myr} of one another, and hence both are shown. }\textbf{Right panel:}  {Probability density distribution of $|\mathbf{\hat{L}}_{\rm sat, infall} \cdot \mathbf{n_{\bot}}|$. The median, shown as a solid line, is equal to 0.56, indicating a preferential infall direction from within the sheet.} }
	\label{fig:subhalo_ang_mom}
\end{figure}

Fig. \ref{fig:subhalo_ang_mom} shows the results in a 2D coloured histogram for the MW and M31 analogues in the LR DMO runs, illustrating that there is a correlation between $|\mathbf{\hat{L}}_{\rm sat,infall} \cdot \mathbf{n}_{\bot}|$ and  $\mathbf{\hat{c}}_{\rm R_{\rm vir}} \cdot \mathbf{n}_{\bot}$. Specifically, it shows that halos with a minor axis currently in the sheet plane (i.e. $\mathbf{\hat{c}}_{\rm R_{\rm vir}} \cdot \mathbf{n}_{\bot} \sim 0$) have had massive mergers that fell in preferentially from outside of the sheet plane (median $|\mathbf{\hat{L}}_{\rm sat,infall} \cdot \mathbf{n_{\bot}}| = 0.33$). On the other hand, halos with a minor axis aligned with $\mathbf{n}_{\bot}$ (i.e. $\mathbf{\hat{c}}_{\rm R_{\rm vir}} \cdot \mathbf{n}_{\bot} \sim 1$) have had massive mergers with a preference for infall from within the sheet (median $|\mathbf{\hat{L}}_{\rm sat,infall} \cdot \mathbf{n}_{\bot}| =  0.69$). Across the full sample of satellite infalls, we find a  {slight preference for }accretion from within the sheet with a median of 0.56,  {as illustrated in the right panel of Fig. \ref{fig:subhalo_ang_mom}}. These results are stable to variations for different numbers of highest mass ratio mergers (one to three), for both the MW and M31, and for both LR and HR DMO runs.  

Focussing now on the hydrodynamical runs, we find that of the three highest mass ratio mergers experienced by our galaxies in the last 12 Gyr of evolution, there is typically one that  {torques} the disk towards its present-day orientation  {\citep[as also seen in e.g.][]{Dillamore2022_ARTEMIS, Bell2026}.} {This is the merger with the largest value of $\mathbf{\hat{L}}_{\rm sat,infall} \cdot~\mathbf{\hat{L}}_{\rm disk}$(today), where a value of one (the maximum) would indicate that the two vectors are fully aligned. We refer to these mergers as disk-defining mergers (DDM) for convenience.  This is the highest mass ratio merger for 6 of the 12 (MW and M31) analogues. In 3 cases, it is the second highest, and in 1 case (MW 1540) it is the third highest. This latter case is a system that falls in and is destroyed a few hundred Myr after the highest mass ratio merger. Finally, in 2 cases, MW 1530 and M31 910, the disk orientation is instead set by the average $\mathbf{\hat{L}}_{\rm sat,infall}$ of two mergers happening in quick succession.} Fig. \ref{fig:subhalo_ang_mom} shows with square symbols the location of such mergers for the MWs, {which follow the trend seen for the  highest mass mergers in the DMO runs.} The star symbols in the figure (whose $y$-coordinates indicate the present-day disk orientation with respect to $\mathbf{n}_{\bot}$, i.e. $|\mathbf{\hat{L}}_{\rm disk} \cdot \mathbf{n}_{\bot}|$)  {follow a similar trend, and we find consistent results for the} M31s (see Appendix~\ref{sec:appendix:M31_mergerdirection}).

\begin{figure}[t!]
	\centering
	\includegraphics[width=\linewidth]{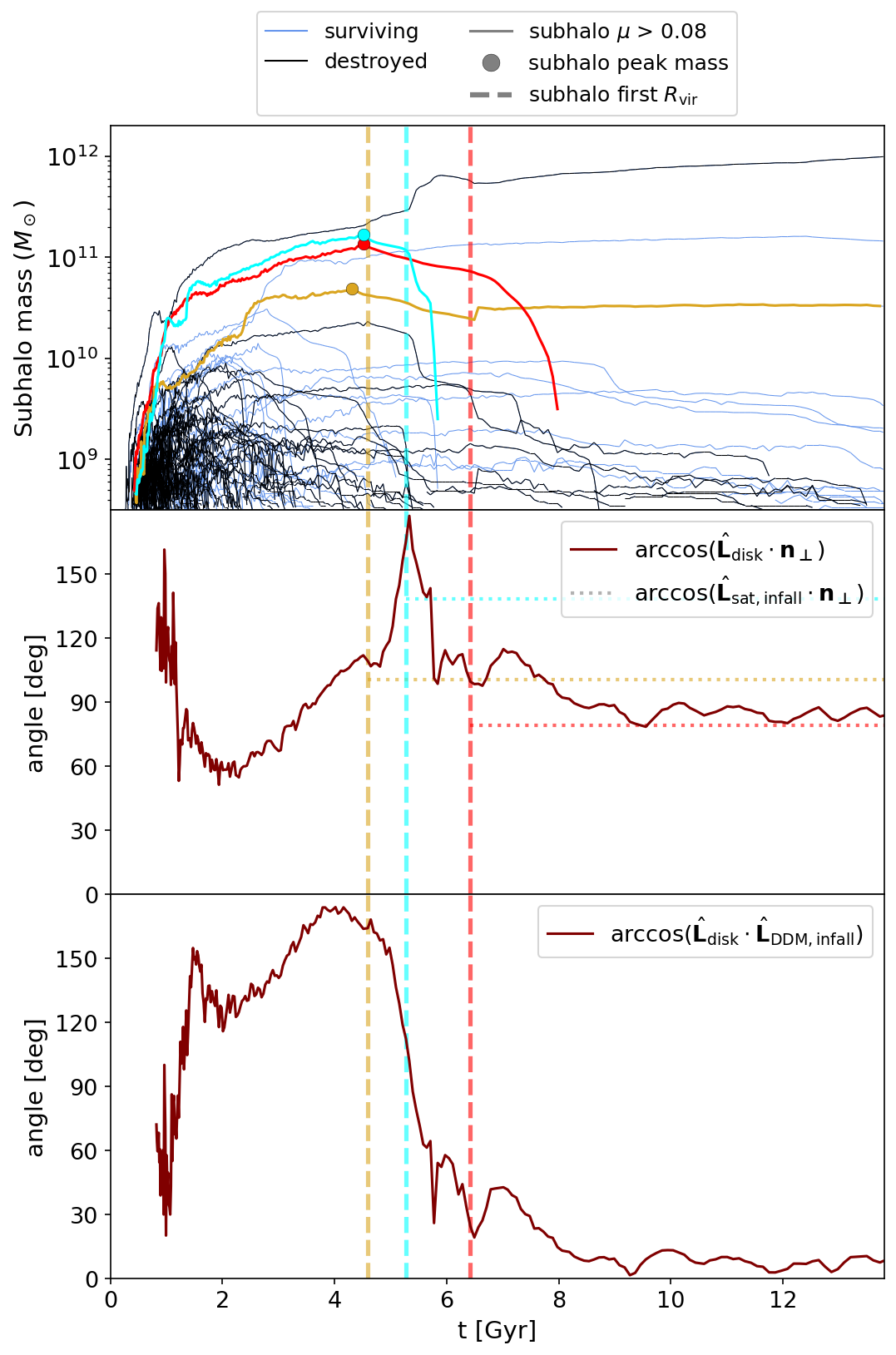} 
	\caption{\small \textbf{Top panel}: Mass growth over time of the MW halo in the LR hydro simulation 720. The lines in light blue show surviving subhalos, while the black lines indicate disrupted subhalos. The thicker lines in different colours indicate subhalos with mass ratios $\mu > 0.08$, the  {round marker} indicates $t_{\rm peak}$, and the dashed vertical lines indicate the infall time ($R_{\rm vir}$ crossing). \textbf{Middle panel}: angle between $\mathbf{n}_{\bot}$ and $\mathbf{\hat{L}}_{\rm disk}$ over time (dark red). Dotted horizontal lines indicate the angle between $\mathbf{\hat{L}}_{\rm sat,infall}$ and $\mathbf{n}_{\bot}$ for each massive merger, colour-coded as in the top panel. \textbf{Bottom panel:} angle between $\mathbf{\hat{L}}_{\rm DDM,infall}$ and $\mathbf{\hat{L}}_{\rm disk}$ over time (dark red).}
	\label{fig:example_infalling_mergers}
\end{figure}

{An example of} the evolution of $\mathbf{\hat{L}}_{\rm disk}$ and of $\mathbf{\hat{L}}_{\rm sat,infall}$ with respect to $\mathbf{n}_{\bot}$  for the three highest mass ratio mergers in the hydrodynamical runs is shown in 
Fig. \ref{fig:example_infalling_mergers} for MW 720. In the top panel the mass growth history is plotted highlighting these massive mergers with different colours (the DDM in red), their peak masses, and times of infall ($R_{\rm vir}$ crossings). The middle panel shows the evolution of the angle $\arccos(\mathbf{\hat{L}}_{\rm disk} \cdot \mathbf{n}_{\bot})$, the dotted lines indicate the angle between $\mathbf{\hat{L}}_{\rm sat,infall}$ and $\mathbf{n}_{\bot}$ at infall time for each merger. A decreasing difference between the two indicates that  $\mathbf{\hat{L}}_{\rm disk}$ reorients itself towards $\mathbf{\hat{L}}_{\rm sat,infall}$. The bottom panel shows the evolution of the angle between $\mathbf{\hat{L}}_{\rm disk}$ and $\mathbf{\hat{L}}_{\rm DDM,infall}$, where small values indicate that the disk and the infall DDM's angular momenta have aligned. This Figure illustrates how $\mathbf{\hat{L}}_{\rm disk}$ first responds to the infall of the cyan merger, after which the subsequent red merger sets the final disk orientation. The infall of a massive subhalo (whose peak mass is similar to the LMC's mass) about at \SI{\sim 13.3}{Gyr} happens too recently to influence $\mathbf{\hat{L}}_{\rm disk}$. We note that typically  {the minor axis direction} $\mathbf{\hat{c}}_{\rm R_{\rm vir}}$ responds faster to the influence of recent infalls, which may lead to misalignments between $\mathbf{\hat{c}}_{\rm R_{\rm vir}} $ and $\mathbf{\hat{L}}_{\rm disk}$ (see also the next section).

\begin{figure}[t!]
	\centering
	\includegraphics[width=\linewidth]{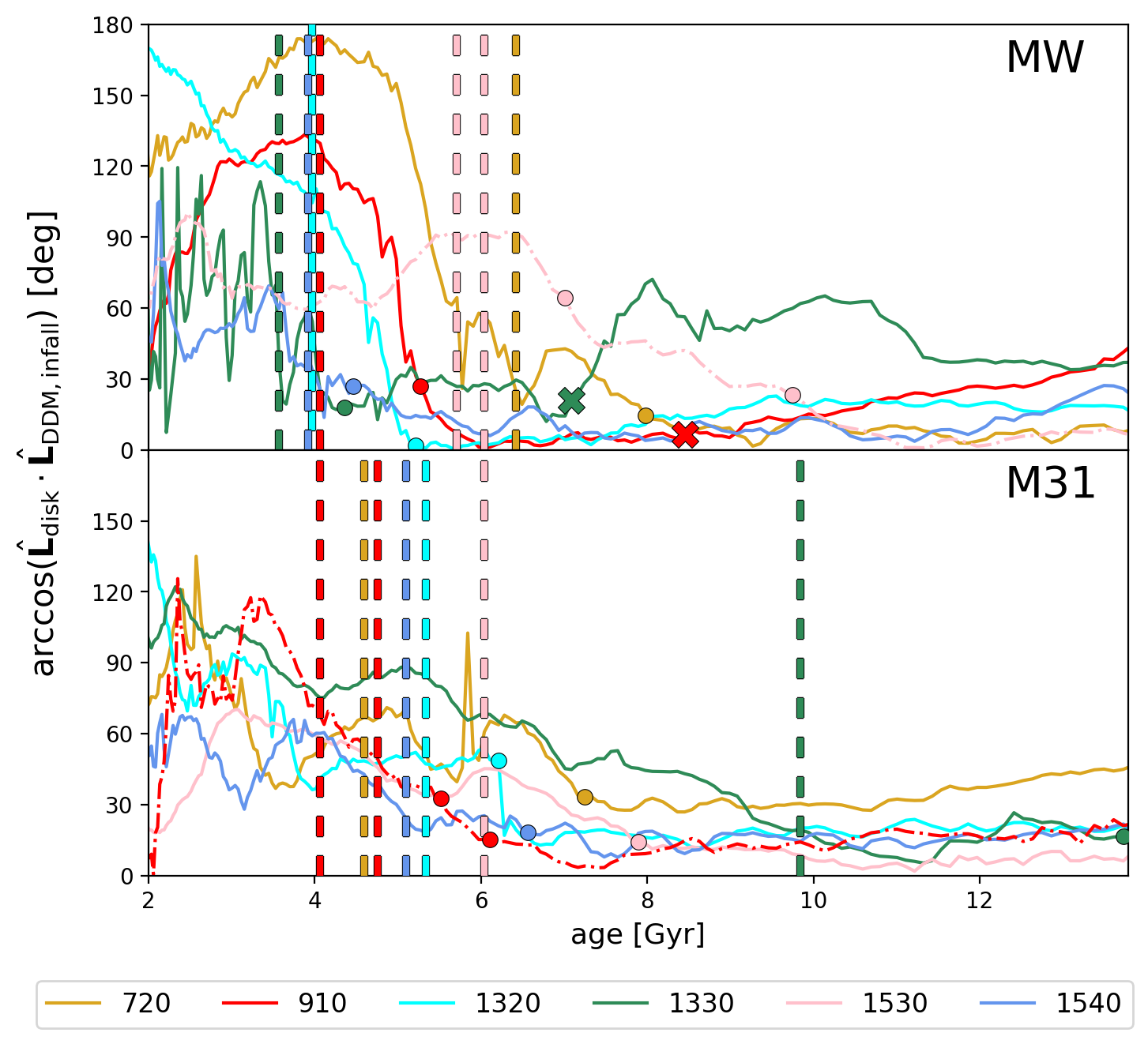} 
	\caption{\small  {Angle between $\mathbf{\hat{L}}_{\rm DDM,infall}$ and $\mathbf{\hat{L}}_{\rm disk}$ over time for the MW (top) and M31 (bottom) analogues in the six LR hydro runs, colour-coded as indicated the legend. Dashed vertical lines indicate the infall time ($R_{\rm vir}$ crossing) of each DDM, filled circles indicate their destruction time. For MW 1530 and M31 910 (dashed-dotted lines), two mergers occurring in quick succession (and with similar angular momenta in the case of MW 1530) jointly set the disk orientation, and both are therefore shown. In these two cases, $\mathbf{\hat{L}}_{\rm DDM,infall}$ is taken as the average of the $\mathbf{\hat{L}}_{\rm sat,infall}$ of both satellites. The crosses for MW 910 and MW 1330 indicate the infall times of satellites on retrograde orbits that subsequently torque the disk. }} 
	\label{fig:DDM_Ldisk_over_time}
\end{figure}

{Similar behaviour is observed for almost all MW and M31 analogues, as illustrated in Fig. \ref{fig:DDM_Ldisk_over_time}. This figure shows the evolution of $\mathbf{\hat{L}}_{\rm DDM,infall} \cdot~\mathbf{\hat{L}}_{\rm disk}$ for all six LR hydrodynamical runs. We see that the disk reorients in response to the infall of the DDM due to the conservation of total angular momentum \citep[e.g.][]{Dodge2023, Bell2026} and generally settles into its new orientation around the time the DDM is destroyed (indicated by the filled circles). Variations in $\mathbf{\hat{L}}_{\rm DDM,infall} \cdot~\mathbf{\hat{L}}_{\rm disk}$ before the infall of the DDM reflect the influence of earlier mergers, sometimes aligning it towards $\mathbf{\hat{L}}_{\rm DDM, infall}$ (e.g. MW 720 or 1320) or away from it (e.g. MW 910 or MW 1530).} On average, the angular momentum of the DDM and that of the present-day disk are aligned within $22 \degree$,  {with 6 (9)  cases showing a misalignment smaller than \SI{20}{\degree} (\SI{27}{\degree}). The 3 cases where the misalignment is larger can be readily be explained. Both MW 910 and MW 1330 experience retrograde mergers with $\mu \sim 0.17$ after the DDM (crosses in Fig.~\ref{fig:DDM_Ldisk_over_time}), which leads to poorer alignment because the orbital and internal spin of the disk are in opposite directions. For M31 720, the DDM is a major merger ($\mu  = 0.9$) implying that the internal spins of both the host and DDM play a role in setting the final orientation.} 

{The average alignment we find is stronger than predicted from random infall orientations. The expected minimum angle between a fixed vector and three randomly drawn unit vectors on a sphere is $54^{+31}_{-27}$~degrees ($47^{+37}_{-27}$~degrees when pairwise mean directions are included). 
 {We find that in general, }the DDM had its first $R_{\rm vir}$ crossing 8-10~Gyr ago, and on average the DDMs fall in 1~Gyr earlier for the MW analogues than for the M31s  \citep[see also][]{Wempe2025a}. M31 1330 is the sole exception to this trend, having experienced a notably recent DDM whose first $R_{\rm vir}$ crossing took place just 4 Gyr ago.}

\section{Discussion} 
\label{sec:disc}

\subsection{Implications for the MW and its observed DM halo and disk alignment}
 
As mentioned in the \nameref{sec:intro} (see also Fig. \ref{fig:cartoon}), the MW's DM halo shape varies radially, transitioning from oblate/near-spherical in the inner regions, with the minor axis aligned with $\mathbf{L}_{\rm disk, MW}$, to prolate/triaxial farther out, with the major axis roughly aligned with $\mathbf{L}_{\rm disk, MW}$. The MW's merger history has been relatively quiescent since its last major merger \SI{\sim 10}{Gyr} ago \citep[e.g.][and references therein]{Helmi2020}. It is currently merging with Sagittarius \citep[Sgr, peak mass \SI{\sim6e10}{\Msun}, e.g.][]{Gibbons2017, Laporte2019}, which likely fell in \SI{\sim 8}{Gyr} ago and reached its first pericenter \SI{\sim 6}{Gyr} ago \citep{RuizLara2020}, and the more massive LMC \citep[peak mass \SI{\sim2e11}{\Msun}][]{Erkal2019, Sheng2024}, which is likely on its first infall \citep{Besla2007, Sheng2024}, although a second infall scenario has also been proposed \citep{Vasiliev2023_LMC}. Since the orbital planes of Sgr and the LMC are roughly perpendicular to the MW's disk plane (and infalling from within the sheet within $\sim 45$\degree), neither is likely to have significantly tilted the MW's disk \citep[see][who showed that neither co-planar nor perpendicular mergers induce tilting]{Dodge2023}. 

\begin{figure}[t!]
	\centering
	\includegraphics[width=0.9\linewidth]{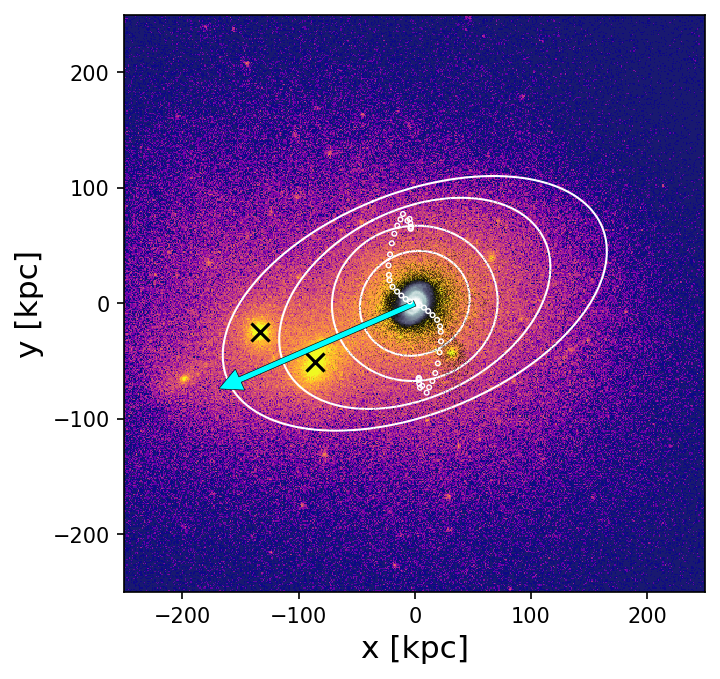} 
	\caption{\small Shape ellipsoids  {projected onto the $(x - y)$ plane }(white) of HR hydro M31 720 at radii of 50, 75, 150, and \SI{200}{kpc}, derived using all DM particles, including those in satellites. White open circles mark the projected direction of the minor axis at radii from 10 to \SI{200}{kpc}c in steps of \SI{10}{kpc}. The cyan arrow indicates the projected direction of $\mathbf{L}_{\rm disk}$, which was set by the DDM \SI{\sim8}{Gyr} ago. DM particles are colorcoded by $\log_{10}$ density (blue to yellow colormap), as are the star particles (black to white colormap). The pair of satellites at a distance of \SI{\sim 100}{kpc} from the centre have masses \SI{\sim1.6e11}{\Msun} and \SI{\sim1.0e11}{\Msun},   {and their centres are indicated with black crosses}. The pair fell in from within the sheet, and as a consequence $\mathbf{\hat{c}}$ has become increasingly perpendicular to this direction (i.e. towards $\mathbf{n}_{\bot}$) at larger radii. }
	\label{fig:incl_excl_sats_small}
\end{figure}

However, the LMC naturally explains the MW's observed prolate outer halo shape and its orientation as a result of infall from within the Local sheet.  {It seems that the} MW's halo  minor axis  {at larger radii} is aligned with $\mathbf{n}_{\bot}$ \citep{Lawmajewski2010,VeraCiro2013,VasilievTango2021}, in agreement with the trend seen in Fig. \ref{fig:minor_NGP_angle}. This means that, {if} another massive merger (and also not Sgr) set the present-day orientation of the MW's disk, {it likely fell} in along $\mathbf{n}_{\bot}$.  {Then, given the MW's quiet merger history, }this leaves  {Gaia-Enceladus-Sausage (GES)} to be the most likely candidate, with its retrograde orbit suggested to have been inclined  by $15-45\degree$ with respect to the MW's disk \citep{Helmi2018,Koppelman2020,Naidu2021}. A good analogue of this scenario is M31 720 shown in Fig. \ref{fig:incl_excl_sats_small}, whose $\mathbf{L}_{\rm disk}$ points into the sheet plane and whose DM halo minor axis, aligned within $30 \degree$ with $\mathbf{L}_{\rm disk}$  at small radii, twists towards $\mathbf{n}_{\bot}$ at large radii due to the infall of a massive pair of satellites \SI{\sim 1.5}{Gyr} ago\footnote{In contrast, MW 720 is not a suitable analogue: although its $\mathbf{L}_{\rm disk}$ aligns with the sheet plane, its recent massive satellite infall (the most massive present-day subhalo in Fig. \ref{fig:example_infalling_mergers}) arrives from outside the sheet and therefore does not drive the minor axis to align with $\mathbf{n}_{\bot}$}.

{Interestingly, in linear theory the $\mathbf{n}_{\bot}$ direction is the direction of maximum collapse for a sheet, and it was likely set $\sim 10-11$ Gyr ago according to our analyses in  Sec.~\ref{sec:sheet-orient}, which is also approximately the estimated time of the merger with GES. Because the MW experienced no other major mergers after GES from \enquote{unfavourable} directions (i.e. leading to tilting of the disk), it is likely its disk has retained the  orientation set at very early times.} 

The mechanism described also provides an explanation for the twisted DM halo shape of the MW, a finding that is supported by recent studies \citep[e.g.][]{VasilievTango2021, Nibauer2025, Dillamore2026}.  
Twisting is not uncommon in cosmological simulations: 32\% of 25 MW-like galaxies in TNG50 exhibit twisted DM halos \citep{Emami2021}. Notably, in our suite of hydrodynamical simulations only M31 720 showed strong twisting up to 80$\degree$, while the other systems show twisting below 30$\degree$.

Moreover, the mechanism presented here may offer  {insights into} the observed plane of satellites of the MW \citep[e.g.][]{Lynden-Bell1976, Kroupa2005}.  If the disk orientation has not changed since the GES merger, 
the accretion of satellites will be preferentially from within the sheet. Interestingly, we find that M31 720's bright satellites preferentially lie in the sheet plane, roughly perpendicular to its disk, {\it and} that they share similar angular momentum orientations due to the infall of the massive satellite pair. These findings are consistent with \cite{Shao2021}, who show that  {for MW-mass halos in the EAGLE simulations} satellite planes tend to be oriented perpendicular to the dark matter halo's minor axis.  {The significance of this configuration of satellites compared to a random distribution can be quantified by the clustering score, defined as $\Delta_{\rm sph}(k) = \sqrt{\frac{1}{k}\sum_{i = 1}^k \left( \arccos( \mathbf{\hat{L}_{\rm mean} \cdot  \mathbf{\hat{L}}}_{i} ) \right)^2}$, where $\mathbf{\hat{L}}_{\rm mean}$ is the unit vector of the mean angular momentum direction of all $k$ satellites, and $\mathbf{\hat{L}}_{i}$ is the unit vector of the angular momentum of the $i$-th satellite \citep[e.g.][]{Pawlowski2013}. The HR M31 720 simulation contains only 10 bright satellites at the present-day},  {and we find }$\Delta_{\rm sph}^{\rm M31 \: 720}(10) = 61 \degree$,  {which is more than $1 \sigma$ below what would be expected from an isotropic distribution}\footnote{{For an isotropic distribution and small $k$, the expected clustering score is not equal to \SI{90}{\degree} but smaller,
i.e. $\Delta_{\rm sph}^{\rm isotropic}(11) = 80.4^{+7.8}_{-9.1}$ degrees, and $\Delta_{\rm sph}^{\rm isotropic}(10) = 79.4^{+8.3}_{-9.5}$ degrees.}}.  {Interestingly, it is consistent (within 1$\sigma$) with the observed clustering score for the MW computed via a bootstrap over 10 of its 11 classical dwarf satellites \citep[using phase-space coordinates from Table 2 in][]{Pawlowski2020}, which is $\Delta_{\rm sph}^{\rm MW}(10) = 55.6 \pm 5.6 \degree$.}

\subsection{Halo shapes, halo spin and minor axis alignment in sheet environments}

We have found that the halo minor axis  {direction} $\mathbf{\hat{c}}_{\rm R_{\rm vir}}$ is  influenced by the local environment, whereas the total angular momentum $\mathbf{J}_{\rm tot}^{\rm R_{\rm vir}}$ shows no significant alignment with $\mathbf{n}_{\bot}$ (i.e. consistent with a uniform orientation). This is in agreement with the work by \cite{Codis2014, Kraljic2020, AragonCalvo2020}, who find a mass-dependent spin flip which implies an absence of a strong alignment signal in the MW and M31-mass regime. 
 {Our results show that $\mathbf{J}_{\rm tot}^{\rm R_{\rm vir}}$ contains geometric information about halo assembly, which can be traced by the observable  $\mathbf{L}_{\rm disk}$  given its strong alignment with $\mathbf{J}_{\rm tot}$ (see Appendix \ref{sec:appendix:convergence:minoraxis_Ldisk}).} 

The misalignment between the angular momentum of DM halos $\mathbf{J_{\rm tot}}$ and their minor axis $\mathbf{\hat{c}}$ has been shown to increase with radius \citep[e.g.][]{Deason2011, Valenzuela2024}, which is confirmed by our findings (see Appendix \ref{sec:appendix:convergence:Jr_minor}). For filament environments, \cite{GaneshaiahVeena2019} further show that  {the} alignment is tighter for galaxies of all masses whose minor axis lies perpendicular to the filament (see their Fig. 13), i.e. perpendicular to the preferential direction of accretion. We observe a comparable trend in the sheet environment studied here: $\mathbf{J}_{\rm tot}^{\rm R_{\rm vir}}$ and $\mathbf{\hat{c}}^{\rm R_{\rm vir}}$ are more closely aligned when either is perpendicular to the sheet.

 {
We find that the halo shapes of the MW and M31, in both the DMO and hydrodynamical runs, are on average rounder than what is reported in the literature. At the virial radius, we find statistically significant differences in the shape parameters $p$ and $q$ relative to the Auriga halos analysed in \citet{Prada2019}. These rounder halo shapes may reflect the influence of the sheet environment, the fact that the two systems form a pair, or the particular choice of constraints used in the initial-condition reconstruction. In particular, our Local Group analogues are drawn from a posterior defined by constraints on the filtered mass {\it and} centre of mass, unlike the comparison samples which are selected based on their present-day $M_{\rm vir}$. This difference in selection may lead to a small bias toward more symmetric mass distributions, potentially contributing to the rounder halo shapes found here. An end-to-end test of the methodology, that sequentially turns on and off the various constraints is a decisive way to test this, but is beyond the scope of this work. 
Next to this, a comparison with isolated halos in sheet environments and with halo pairs in random environments would also be desirable but no suitable literature study exists to-date to the best of our knowledge.}

\section{Conclusions}
\label{sec:conclusion}

We have studied the dark matter halo shapes and (disk) orientations of MW and M31 analogues with respect to their local environment using the constrained Local Group simulations described in \cite{Wempe2024, Wempe2025a, Wempe2026} and \cite{Geninainprep}. In  {all of these simulations, the Local Group is embedded within a sheet-like environment,} analogous  {and almost parallel to the observed} Local Sheet.  {We describe the orientation of the simulated Local Sheets by their} normal vector $\mathbf{n}_{\bot}$. We have run convergence tests in our 
statistical sample of DMO and  {in our} smaller set of hydrodynamical simulations at various resolutions, and find good agreement for all quantities of interest, including  the DM halos' minor axis direction  and total angular momentum. 

The DMO halos in our simulations are on average triaxial, with the MWs being slightly rounder than the M31s, though consistent within  {the} scatter. As expected, the hydrodynamical runs produce rounder DM halos, with $q \sim 1$ in the inner regions where baryonic condensation drives the shape towards being on average oblate. Farther out, the halo shapes tend to become triaxial or near triaxial, consistent with the generally radially varying shapes reported in the literature  \citep{Zavala2019, Shao2021}. However, compared to the literature our halos are  {on average} slightly rounder  {than typical}, perhaps reflecting the fact that the MW and M31 are a galaxy pair or that they are embedded in a particular environment.

In terms of alignments, we find that the disk angular momentum is generally closely aligned with  {the DM halo minor axis direction }(median alignment within 15$^{+15 }_{-8 }$  {degrees} at $R_{\rm vir}$) and  {with the DM halo total angular momentum (median alignment within $25^{+20}_{-16}$  {degrees} at $R_{\rm vir}$)}. We find that massive mergers ($M_{\rm halo} \geq$ \SI{e11}{\Msun}) can induce misalignments of many tens of degrees {, and} that the sheet environment causes  alignment between the  {minor axis direction at the virial radius} and $\mathbf{n}_{\bot}$, consistent with  previous work \citep[e.g.][]{AragonCalvo2007, Codis2018}, with a median  {cosine} alignment of $\sim 0.7$ across the DMO runs. We find no significant correlation between  {the total DM halo angular momentum at the virial radius }and $\mathbf{n}_{\bot}$. Using the DMO simulations, we find that the infall direction of massive mergers (mass ratio $\mu > 0.08$)  influences the orientation of  {the minor axis direction at the virial radius}, such that halos with  {a minor axis}  {parallel to the sheet plane }are more likely to have experienced  {massive} mergers originating from outside of the sheet, while those  with  {a minor axis}  {aligned with the sheet normal} are more likely to have accreted massive satellites from within the sheet. For the full sample of massive mergers, we find preferential accretion  {from directions }within the sheet.

We have also investigated how massive mergers affect the present-day disk orientation. We find that the disk orientation is  {often} set by one of the two highest mass ratio mergers experienced in the last {8 -- 10} Gyr, and typically the most massive one, although this depends slightly on the timing of the merger itself. Indeed, we find that the angular momentum of such a  merger and that of the present-day disk are typically aligned within $22\degree$.
While recent satellite infall  {events} ($\lesssim 2$~Gyr) can reorient the halo's minor axis at the virial radius,  {the disk angular momentum} generally does not have sufficient time to reorient in response to these late merger events, keeping an imprint of earlier significant mergers.

 {Studies in the literature seem to support a picture in w}hich the MW's elongated DM halo at \SI{\sim 50}{kpc} has its minor axis aligned with the normal to the Local Sheet, while its disk angular momentum is aligned with the plane of the Local Sheet.  {This can be understood in} the context of our results,  {which demonstrate that disk orientations may be set by the infall direction of massive mergers}. Since Sgr and the LMC have both fallen in from within the sheet and their orbital plane is almost perpendicular to the MW's disk direction, neither seems to be responsible for the current disk orientation.  {Given the MW's quiescent merger history, } this leaves GES as the most likely candidate to have set the disk's current orientation \citep[the inclination of its orbit has been estimated to be $\sim 15-45\degree$,][]{Helmi2018,Naidu2021},  {with its infall direction broadly coinciding with the direction of maximum collapse of the Sheet.} Furthermore, the prolate shape of the MW's DM halo can be explained by the LMC's infall \citep[as also seen by][]{VeraCiro2013, VasilievTango2021}, driving the minor axis to align with $\mathbf{n}_{\bot}$, resulting in a twisted DM halo. M31 720 presents an example of this mechanism, as the merger that defined the disk orientation came from outside of the sheet, while the current infall of a pair of galaxies (peak mass similar to the LMC) from within the sheet twists the halo minor axis towards $\mathbf{n_{\bot}}$. Intriguingly, the satellites of M31 720 organize themselves in a plane perpendicular to the disk, with a degree of clustering  {that is significantly different from a random configuration, and similar to that observed for the MW satellites.}

Our work demonstrates  {the value of statistically large samples} of realistic constrained cosmological simulations  {for connecting galaxy properties to their formation context}. {The orientation of the disks of the MW and M31 are at most weakly related to the large-scale sheet environment and so clearly contain further significant information about galaxy assembly. Since we find that the disk angular momenta align well with the angular momenta of their DM halos, the latter could be used as additional constraints to obtain even higher fidelity simulations of Local Group assembly. However, reproducing the twisted DM halo of the MW may be more challenging.} 

\begin{acknowledgements}
This research has been supported by a Spinoza Grant from the Dutch Research Council (NWO, SPI78-411). HCW thanks Francesco Guarneri for helpful discussions, and acknowledges the help of LLM in writing this manuscript and specific coding tasks. AG is supported by UK Research and Innovation (UKRI) under the UK government’s Horizon Europe funding guarantee [grant number EP/Z534353/1]. Part of this research was carried out using the High-Performance Computing resources of the FREYA cluster at the Max Planck Computing and Data Facility (MPCDF, https://www.mpcdf.mpg.de) in Garching, operated by the Max Planck Society (MPG).

Throughout this work, we have made use of the following packages: \texttt{astropy} \citep{Astropy},
         \texttt{matplotlib} \citep{matplotlib},
         \texttt{NumPy} \citep{Numpy},
         \texttt{GADGET-4} \citep{Springel2021},
         \texttt{H5PY},
         \texttt{SciPy} \citep{2020SciPy-NMeth},
         \texttt{tqdm} \citep{tqdm},
         and Jupyter Notebooks \citep{JupyterNotebook}. 

Finally, we acknowledge the use of \texttt{WebPlotDigitizer} \citep{plotdigitizer}.
\end{acknowledgements}

\bibliography{bibliography}
\bibliographystyle{aa} 

\onecolumn

\begin{appendix}
\label{sec:appendix}

\section{q, s profiles in DMO and hydrodynamical runs} 
\label{sec:appendix:qsprofiles}

\begin{figure}
	\centering
	\begin{subfigure}{0.76\textwidth}
		\centering
		\includegraphics[width=\linewidth]{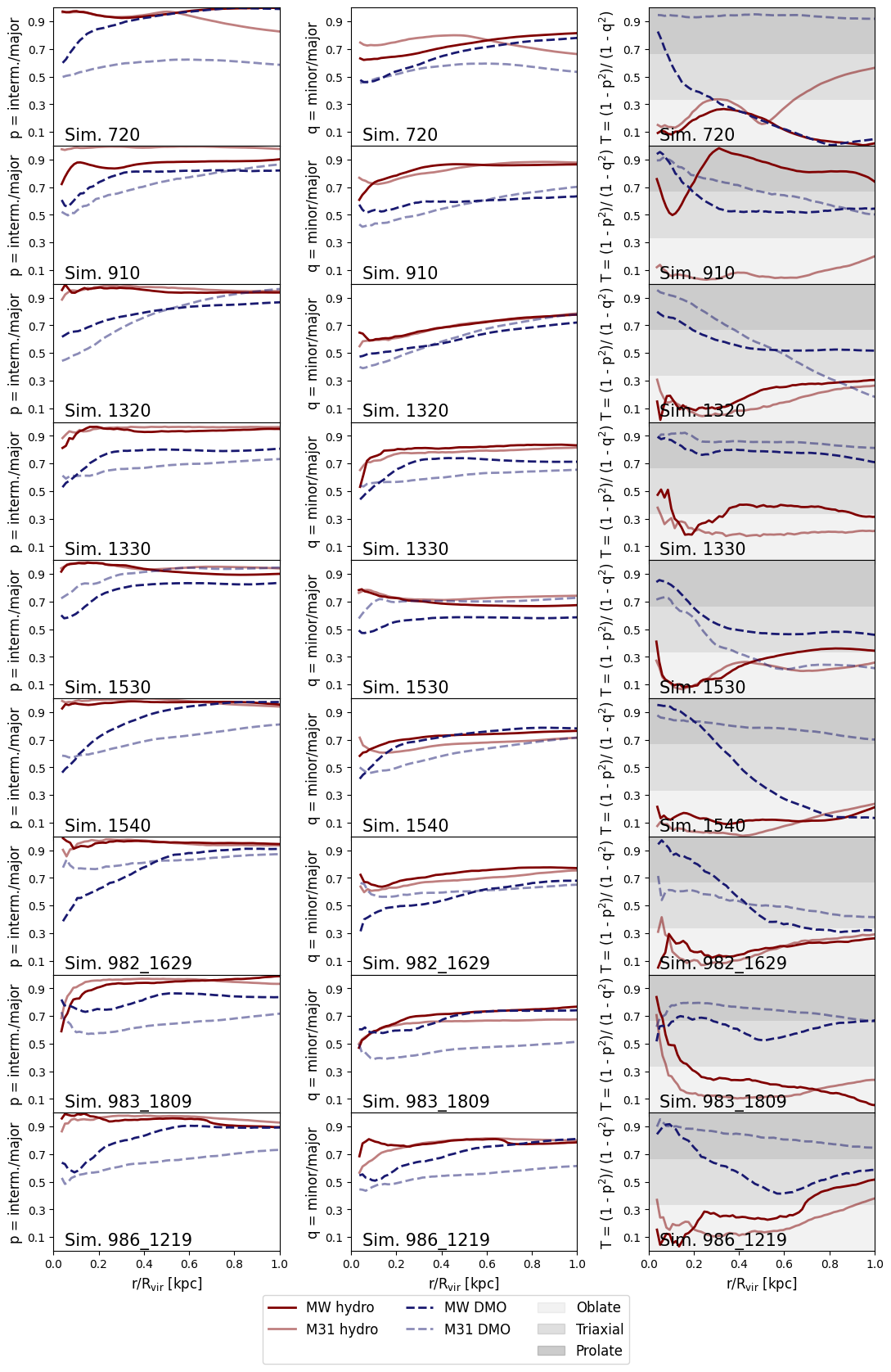} 
	\end{subfigure}
	\caption{\small   Shape profiles as a function of radius for the MW and M31 (the latter shown in lower opacity) for all hydrodynamical runs (brown solid lines), compared to  their corresponding DMO runs (dark blue dotted lines) using the highest resolution available. Results are shown for the highest resolution available for each run.  \textbf{Left column:} Ratio of the intermediate over the major axis $p$ as a function of radius. \textbf{Middle column:} Ratio of the minor over the major axis $q$ as a function of radius . \textbf{Right column:} Triaxiality, defined as $T = (1 - p^2)/(1 - q^2)$, as a function of radius. The differently grey shaded areas indicate $0 \leq T \leq 1/3$, corresponding to oblate shapes, $1/3 \leq T \leq 2/3$, corresponding to triaxial shapes, and $2/3 \leq T \leq 1$, corresponding to prolate shapes. Most halos are oblate, see Fig. \ref{fig:shapes_MW_M31_Rvir}. MW 910 is an exception, deviating from this trend due to the influence of a merger remnant of an LMC-SMC like pair of subhalos around $r$  \SI{\sim 50}{kpc}. Similarly, M31 720 tends toward a nearly prolate shape in its outer regions due to the recent infall of a massive satellite pair. Compared to the DMO runs, which are generally triaxial, the hydrodynamical runs are more axisymmetric ($p \sim 1)$  and oblate as a result of baryonic condensation.}
	\label{fig:T_profiles}
\end{figure}

Fig. \ref{fig:T_profiles} shows a comparison of $p$ \textbf{(left column)} and $q$ \textbf{(middle column)} as a function of radius for the LR hydro, HR hydro, LR DMO and HR DMO runs using the highest resolution available.  {The right column of} Fig. \ref{fig:T_profiles} shows the triaxiality, which is defined as $T = (1 - p^2)/(1 - q^2)$, where $0 \leq T \leq 1/3$ corresponds to oblate shapes, $1/3 \leq T \leq 2/3$ corresponds  to triaxial shapes, and $2/3 \leq T \leq 1$ corresponds to prolate shapes. 
The hydrodynamical runs are rounder and  {more} oblate (close to being axisymmetric with $p$ approaching unity) in the inner regions, while in the outer regions they tend to become triaxial. This is consistent with literature \citep[e.g.][]{Chua2019}, although our halos appear to be slightly rounder.  We notice that the effect of baryons is to turn the halo shapes, which are often prolate in the centre in the DMO runs, to an oblate configuration, and while the outer shape is also changed, it is not  {changed} as dramatically. 

\section{Convergence and correlations between minor axis direction, angular momentum and $\mathbf{L}_{\rm disk}$} 
\label{sec:appendix:convergence}

\subsection{Convergence in the directions of the halo minor axis and $\mathbf{J}_{\rm tot}(r)$ in DMO and hydrodynamical runs}

\begin{figure*}
	\centering
    \vspace{-3pt}
    	\begin{subfigure}{\textwidth}
		\centering
		\includegraphics[width=0.77\linewidth]{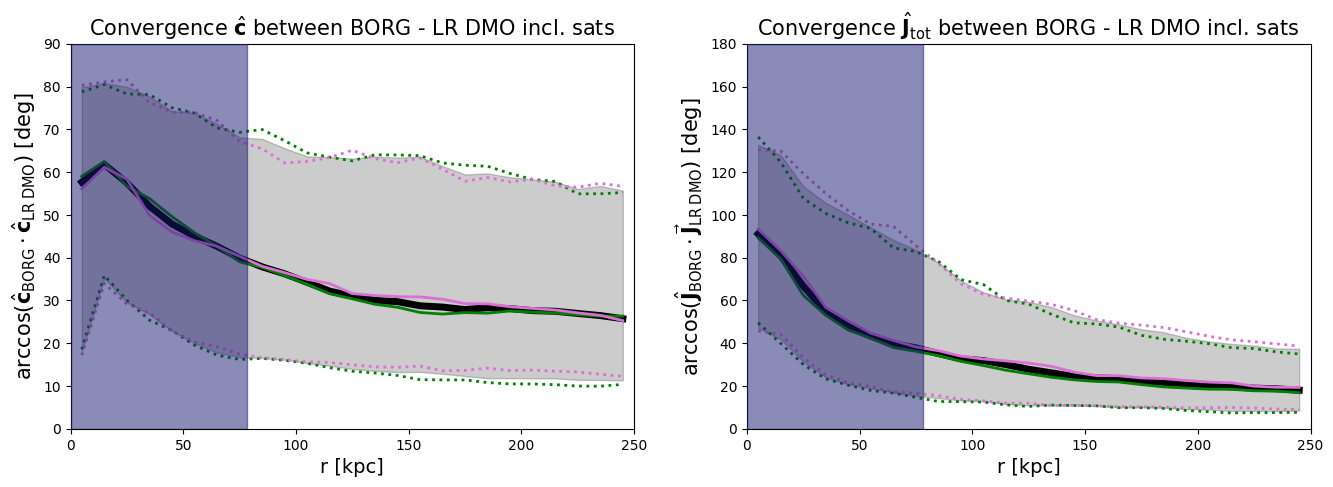} 
	\end{subfigure}
    \vspace{-3pt}
	\begin{subfigure}{\textwidth}
		\centering
		\includegraphics[width=0.77\linewidth]{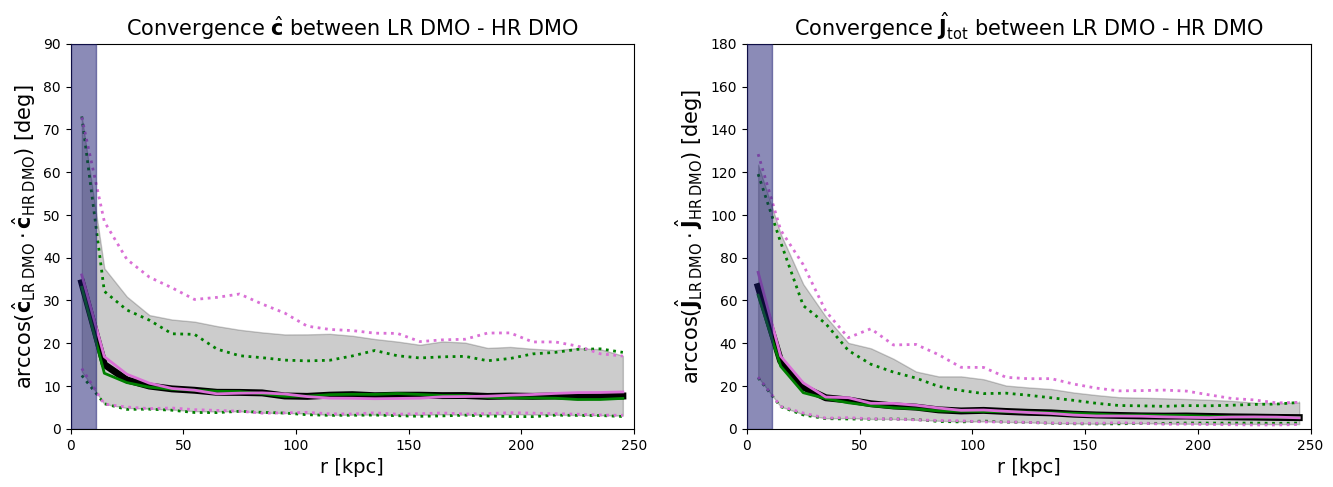} 
	\end{subfigure}
    \begin{subfigure}{\textwidth}
		\centering
		\includegraphics[width=0.77\linewidth]{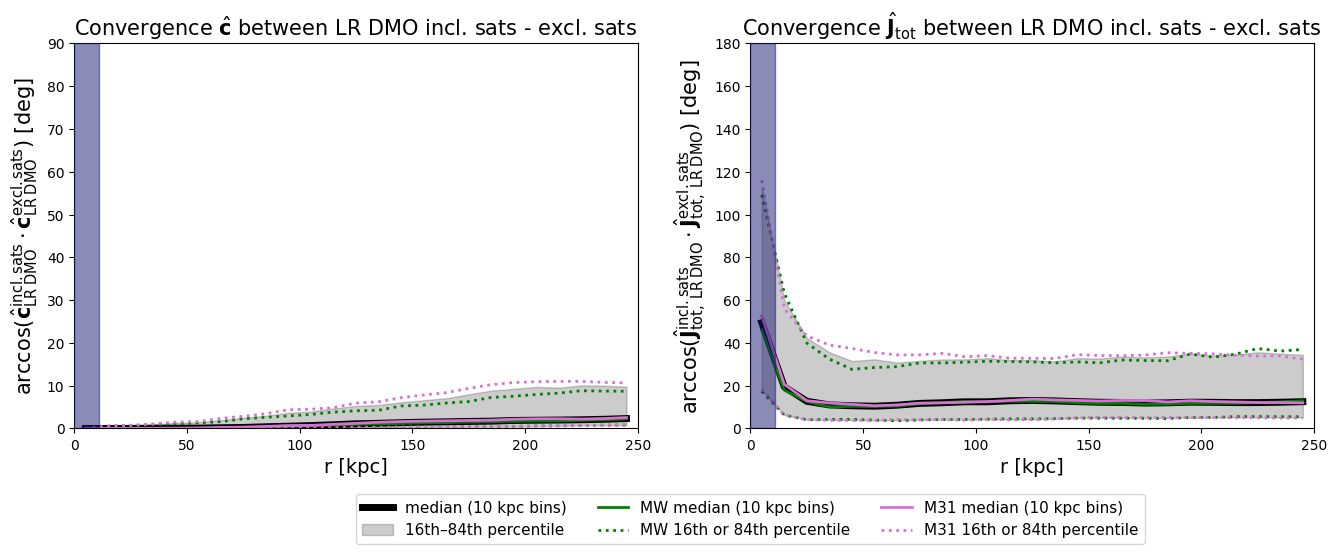} 
	\end{subfigure}
    \vspace{-3pt}
	\begin{subfigure}{\textwidth}
		\centering
		\includegraphics[width=0.77\linewidth]{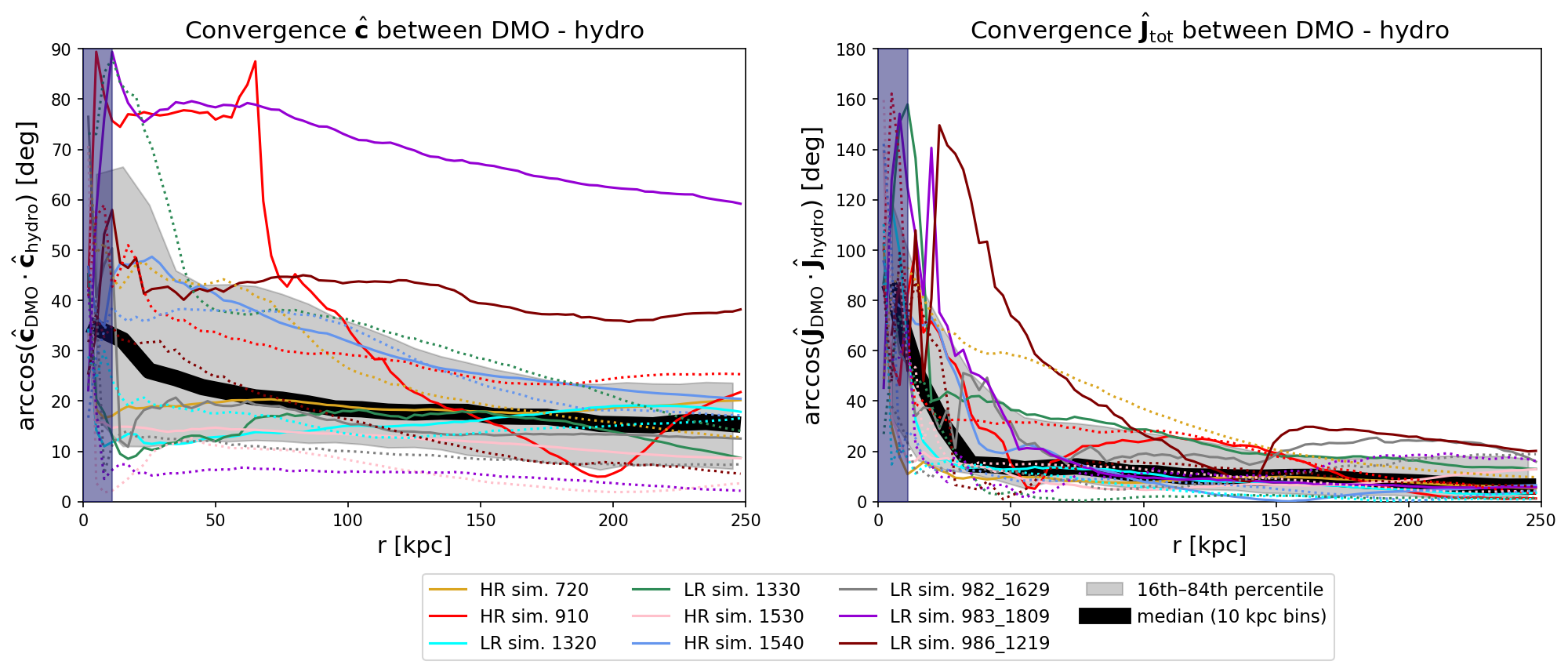}
	\end{subfigure}
	\caption{\small Convergence tests across different resolutions and frameworks for the present day halo minor axis direction (left column) and total angular momentum (right column) as a function of radius. The black line indicates the median, the grey shaded area encompasses the 16$^{th}$ - 84$^{th}$ percentiles, and the shaded dark blue region indicates the inner region where quantities  cannot be reliably determined due to resolution. In the top three rows, solid and dotted green and pink lines show the convergence for the MW and M31 samples separately. \textbf{Top row:}  Convergence between BORG and LR DMO runs using all particles within the volume.
    \textbf{Second row:}  Convergence between HR DMO and LR DMO runs. \textbf{Third row:} Convergence between LR DMO using main halo particles and LR DMO using all particles within the volume. \textbf{Bottom row:} Convergence between DMO and hydrodynamical runs for the MWs (solid lines) and M31s (dotted lines), with each run shown in a different colour (see legend).  {We increased the thickness of the line indicating the median for clarity.}}
	\label{fig:convergence_plots}
\end{figure*}

We first tested how the orientations and directions of the angular momenta of the dark matter halos in the \texttt{BORG} runs compare to the LR DMO simulations. The top row of Fig. \ref{fig:convergence_plots} shows that the directions of the minor axes (determined including the satellites, as the resolution of
the \texttt{BORG} runs is too low to resolve these) agree in the median within 26$^{+30 }_{-15 }$  {degrees} at $R_{\rm vir}$. Their angular momenta directions at $R_{\rm vir}$ are better aligned (within 18$^{+19}_{-10 }$  {degrees}). 
 
We also assessed the convergence between the orientations inferred in the LR and HR DMO runs by computing the angle between the DM halo minor axis directions and the angular momenta $\mathbf{J}_{\rm tot}$ as a function of radius. The second row of Fig. \ref{fig:convergence_plots} shows that the directions of the minor axes for the two resolutions agree on average within 8$^{+9}_{-5}$  {degrees}, while the angular momenta agree within 5$^{+7}_{-3 }$  {degrees}.

We also investigated  the impact of substructure on the shape and angular momentum determination. To quantify their impact on the derived shape, we compute the misalignment between the minor axis determined from particles  {that are }part of the main halo alone, $\mathbf{\hat{c}}_{\rm excl. sats}$, and that determined from all particles in the volume,  $\mathbf{\hat{c}}_{\rm incl. sats}$. The third row of Fig. \ref{fig:convergence_plots} shows that in the inner $\lesssim 100$~kpc, the differences are minimal, in agreement with e.g. \cite{Chua2019, Emami2021}. By inspecting in detail the differences for a number of cases, we find that at larger radii the effect naturally depends on the subhalo's mass. For subhalos with masses  \SI{\sim e10}{\Msun}, the influence is negligible, with  misalignments of only a few degrees. For subhalos with masses of $\gtrsim$  \SI{e11}{\Msun}, the effect instead becomes significant (see for example Fig. \ref{fig:incl_excl_sats_small}), particularly shortly after infall when the satellite has not lost much mass yet, leading to misalignments up to 30$\degree$. However, for a satellite with a peak mass of around  \SI{e11}{\Msun}  that has already lost a significant amount of mass and has sunk into the halo, the misalignment reduces to typically 10$\degree$, as then the halo has also responded to the infall. This is interesting, as it implies that observationally the orientation of the MW's DM halo should be relatively robust, taking into account that the LMC has likely lost a significant amount of mass.

Next, we assessed the convergence between DMO and hydrodynamical runs. The bottom row of Fig. \ref{fig:convergence_plots} shows that, on average, at large radii ($\sim R_{\rm vir}$) the minor axes orientations agree within 15$^{+9}_{-8 }$  {degrees}, while the angular momenta of the DM halos are aligned within 6$^{+11}_{-4}$  {degrees}. Toward smaller radii, the median misalignment of the minor axis increases to \SI{\sim 30}{\degree} while for the angular momentum it increases to \SI{\sim 40}{\degree}, which is due to the expected influence of the baryons in the centre. In a few cases, the misalignment is significant. This can be explained by the fact that in the hydrodynamical runs satellites tend to sink to the centre of the halo faster, meaning mergers can happen on shorter timescales. For example, in MW 910, a massive LMC-like merger remnant is found at ~50 kpc in the hydrodynamical run, while it is farther out in the DMO run. Similarly, \texttt{983\_1809} has fully merged with a massive LMC-like object in the hydrodynamical case, while in DMO it only just passed first pericenter (at \SI{\sim 100}{kpc}), and in \texttt{986\_1219} a satellite pair has already merged in hydro but is just reaching first pericenter (at \SI{\sim 150}{kpc}) in DMO.

\subsection{DMO runs: correlation between $\mathbf{J}_{\rm tot}(< r)$ and the minor axis direction} \label{sec:appendix:convergence:Jr_minor}

\begin{figure}[hbt!]
	\centering
	\includegraphics[width=0.5\linewidth]{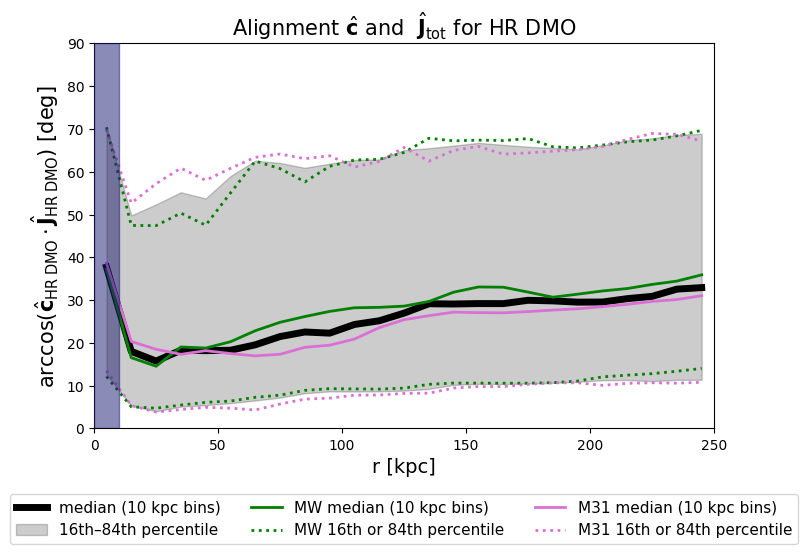} 
	\caption{\small Angle between the present-day halo minor axis direction and $\mathbf{J}_{\rm tot}(<r)$ in HR DMO for all MWs and M31s as a function of radius. The black line indicates the median, the grey shaded area encompasses the 16$^{th}$ - 84$^{th}$ percentiles, and the shaded dark blue region indicates the inner region where quantities  cannot be reliably determined due to resolution. The solid and dotted green and pink lines show this just for the sample of MWs and M31s, respectively. The two quantities are clearly correlated {  but there is a large scatter.}}
	\label{fig:dot_minor_J_DMO}
\end{figure}

We also inspected the angle between the halos' angular momenta and minor axis orientation as a function of radius for the DMO runs. We find that the two quantities are correlated and align within 20$^{+30}_{-15 }$  {degrees} in the inner regions, with misalignment increasing to 30$^{+30}_{-20 }$  {degrees} around the virial radius, as shown in Fig. \ref{fig:dot_minor_J_DMO} for the HR case (consistent results are obtained for the LR and HR DMO runs). This agrees with previous findings, for example by \citet[][their Fig. 2]{Deason2011} or \citet[][their Fig. 16]{Valenzuela2024}.

\subsection{Hydrodynamical runs: correlations between $\mathbf{L}_{\rm disk}$ and the DM halo minor axis direction and its angular momentum}\label{sec:appendix:convergence:minoraxis_Ldisk}

\begin{figure*}[hbt!]
	\centering
	\includegraphics[width=0.9\linewidth]{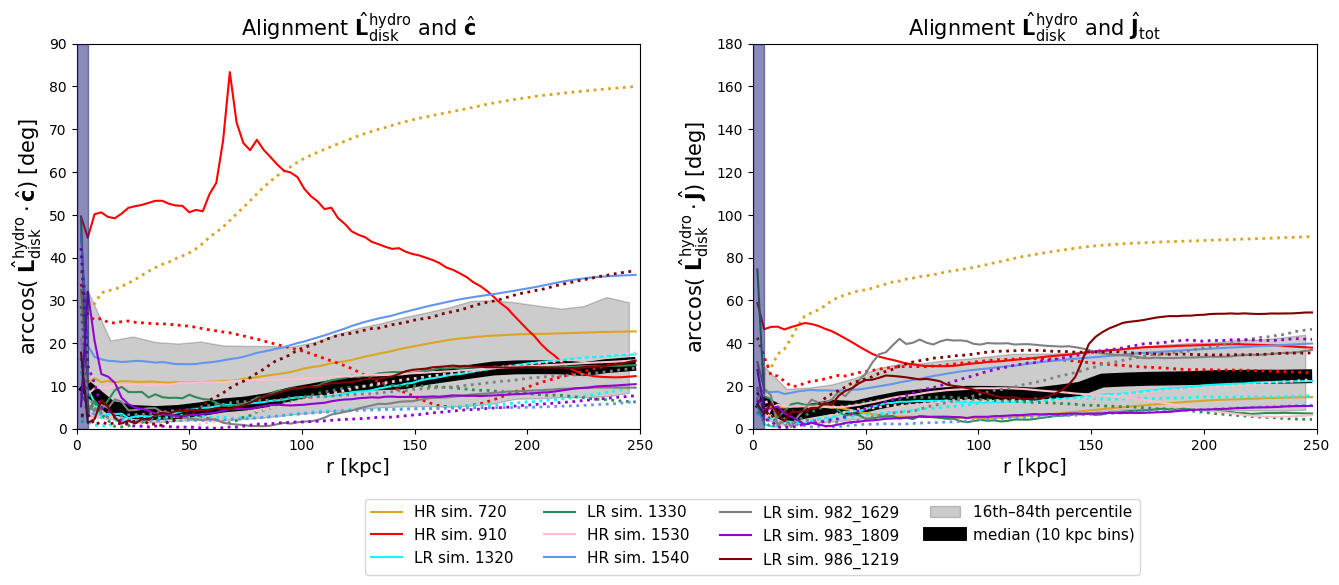} 
	\caption{\small \textbf{Left panel:} Angle between the present-day $\mathbf{L}_{\rm disk}$ and the halo minor axis direction as a function of radius for the MWs (solid lines) and M31s (dotted lines) of the hydrodynamical runs (each run is indicated in a different colour, see legend). The black line indicates the median, the grey shaded area encompasses the 16$^{th}$ - 84$^{th}$ percentiles, and the shaded dark blue region indicates the inner region where quantities  cannot be reliably determined due to resolution.  MW 910 and M31 720 deviate from this trend as they have undergone a recent massive merger, which twists the DM halo minor axis away from the disk plane. \textbf{Right panel:} As the left panel, but showing the angle between the present-day $\mathbf{J}_{\rm tot}(<r)$ and $\mathbf{L}_{\rm disk}$. Also here, M31 720 strongly deviates from the general trend due to the infall of a massive satellite pair. While it is aligned within 30$\degree$ with the disk in the inner region, it becomes misaligned by almost $90\degree$ towards the virial radius. MW 910 and MW 986\_1219 are misaligned due to a merger remnant and a recently merger subhalo, respectively.}
	\label{fig:disk_ang_mom_minor_axis}
\end{figure*}

In the literature, $\mathbf{L}_{\rm disk}$ is generally found to align with the minor axis direction of the galaxy's DM halo, with stronger alignment in their inner regions, as in the outer regions baryonic condensation has less of an effect and the DM halo can twist \citep[see e.g. Fig. 6 in][]{Valenzuela2024}. Here, we investigate for our suite of hydrodynamical simulations how strong this alignment is as a function of radius. The left panel of Fig. \ref{fig:disk_ang_mom_minor_axis} shows the median misalignment is 9$^{+10}_{-7}$  {degrees} at 100 kpc, and increases to about 15$^{+15}_{-8}$  {degrees} at 250 kpc with a somewhat larger scatter. In two cases, for MW 910 and M31 720, there is a large deviation from the trend, which is due to a  {completed merger} and  {due to} a recently merged subhalo pair, respectively. In conclusion, there is good alignment unless the halo has undergone a very recent merger that results in a twisting of the DM halo. 

The right panel of Fig. \ref{fig:disk_ang_mom_minor_axis} also shows the alignment between  $\mathbf{L}_{\rm disk}$ and the total angular momentum of the halo $\mathbf{J}_{\rm tot}(r)$ as a function of radius. The median misalignment is 17$^{+17}_{-11}$   {degrees} at 100 kpc, and increases to 25$^{+20}_{-16}$  {degrees} at 250 kpc with a larger scatter.  {Given that the possible misalignment range is twice as large (i.e. maximum 180$\degree$ instead of 90$\degree$), }this indicates that the alignment is { slightly better} than that with the minor axis direction. 

\FloatBarrier

\section{DDM and disk orientation for M31 hydrodynamical runs}\label{sec:appendix:M31_mergerdirection}

\begin{figure}[t!]
	\centering
	\includegraphics[width=0.5\linewidth]{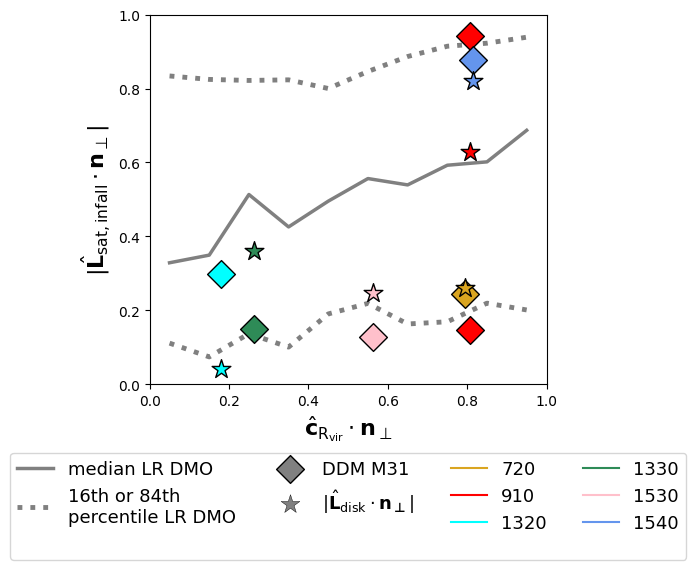} 
	\caption{\small Same as Fig. \ref{fig:subhalo_ang_mom}, but showing the M31s in LR hydrodynamical runs. The $x$-axis represents $\mathbf{\hat{c}}_{\rm R_{\rm vir}} \cdot \mathbf{n_{\bot}}$, the $y-$ axis represents the angular momentum direction at the moment of infall $\mathbf{\hat{L}}_{\rm sat, infall}$ with respect to $\mathbf{n}_{\bot}$ of M31's DDM with $\mu \geq 0.08$. When $|\mathbf{\hat{L}}_{\rm sat, infall} \cdot \mathbf{n_{\bot}}|$=1, the satellite fell in from within the plane of the sheet, while a value of 0 corresponds to infall perpendicular to it. The solid grey and dotted grey lines show the same median and percentiles as in Fig. \ref{fig:subhalo_ang_mom}.   { The $y$-axis coordinate of the stars indicates} the orientation of the disk with respect to $\mathbf{n}_{\bot}$ at the present day, i.e. $\mathbf{\hat{L}}_{\rm disk} \cdot \mathbf{n}_{\bot}$. In the case of M31 910, the averaged influence of two coinciding mergers determines the disk orientation and are thus both included. }
	\label{fig:mergers_hydro_infall_direction_M31}
\end{figure}

The square symbols in Fig. \ref{fig:mergers_hydro_infall_direction_M31} show the distribution of the DDMs' angular momenta at infall and the minor axis of the host halo, both with respect to the direction perpendicular to the sheet $\pm  \mathbf{n}_{\bot}$, for the M31s (i.e. as in Fig. \ref{fig:subhalo_ang_mom}), while the $y$-axis of the star symbols shows the orientation of the disk with respect to $\pm \mathbf{n}_{\bot}$. Most systems follow the trend seen for the LR DMO runs (shown in grey), meaning that a halo whose minor axis is perpendicular to $\mathbf{n}_{\bot}$ (meaning $\mathbf{\hat{c}}_{\rm R_{\rm vir}} \cdot \mathbf{n}_{\bot} = 0$) has mainly undergone mergers coming in from outside of the sheet. For M31 910, the average influence of two mergers ($\mu = 0.2$ and $\mu = 0.3$) that followed each other in quick succession sets the present-day disk orientation, which is why both are included in the Figure. As is the case for the MWs, the disk orientation tends to align with the minor axis direction. 

However, a clear exception is M31 720, whose minor axis is roughly aligned with $\mathbf{n}_{\bot}$, but its $\mathbf{L}_{\rm disk}$ is almost perpendicular to $\pm \mathbf{n}_{\bot}$. This is because M31 720 is currently undergoing a massive merger with an LMC-SMC like pair of galaxies coming in from the sheet, which sets the current minor axis direction and twists the halo (see Fig. \ref{fig:incl_excl_sats_small}). Its disk orientation, however, was set about 8~Gyr ago by a major merger ($\mu = 0.9$). In between these two last major mergers and the recent LMC-SMC-like merger event, no other major mergers took place.

\section{Hydro MW and M31 in LSS} \label{sec:appendix:MW_M31_LSS}

\begin{figure}
	\centering
	\begin{subfigure}{0.97\textwidth}
		\centering
		\includegraphics[width=\linewidth]{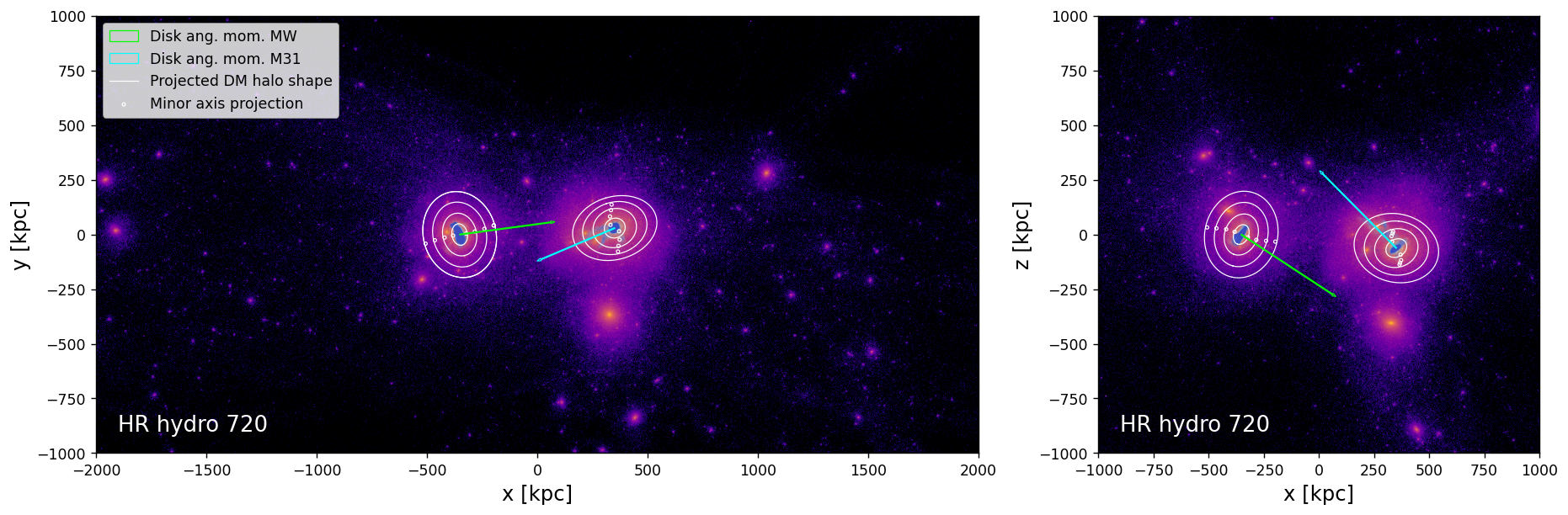} 
	\end{subfigure}\vspace{-0.4em}
	\begin{subfigure}{0.97\textwidth}
		\centering
		\includegraphics[width=\linewidth]{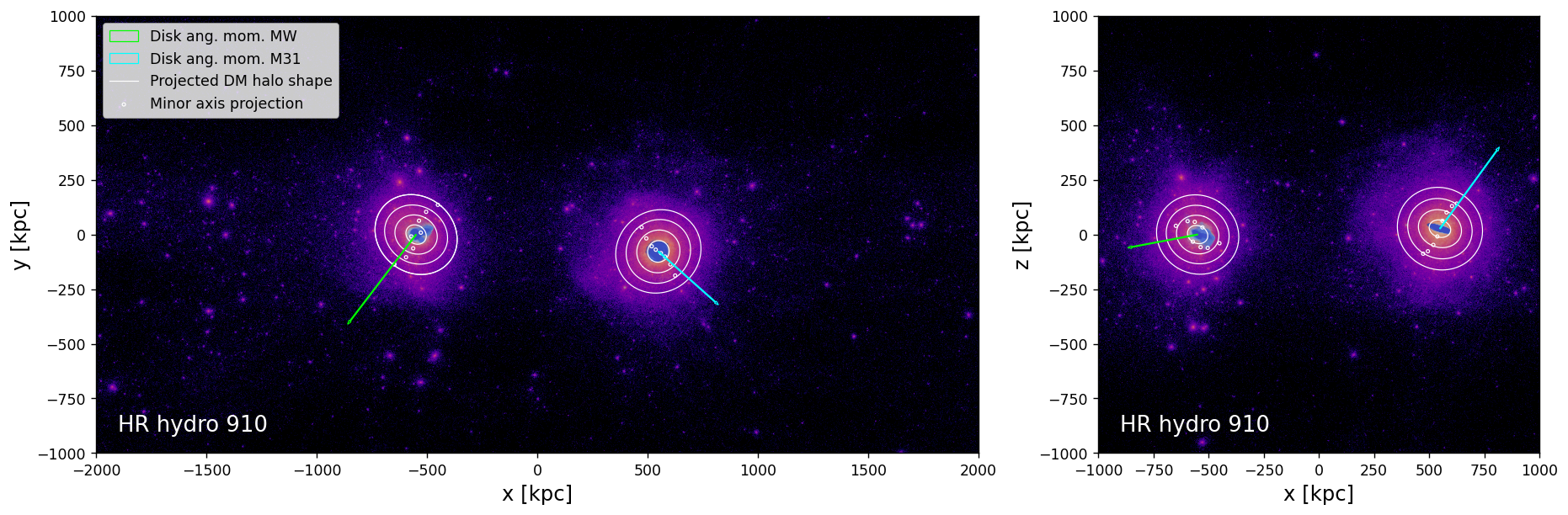}
	\end{subfigure}\vspace{-0.4em}
	\begin{subfigure}{0.97\textwidth}
		\centering
		\includegraphics[width=\linewidth]{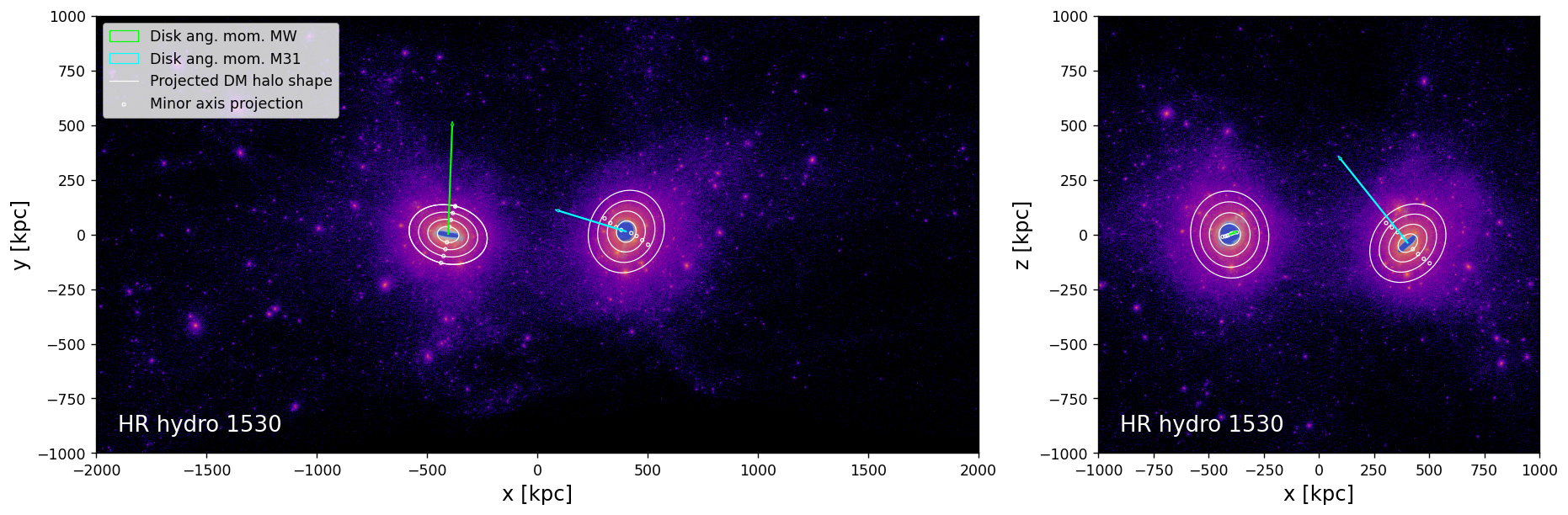}
	\end{subfigure}\vspace{-0.4em}
    \begin{subfigure}{0.97\textwidth}
		\centering
		\includegraphics[width=\linewidth]{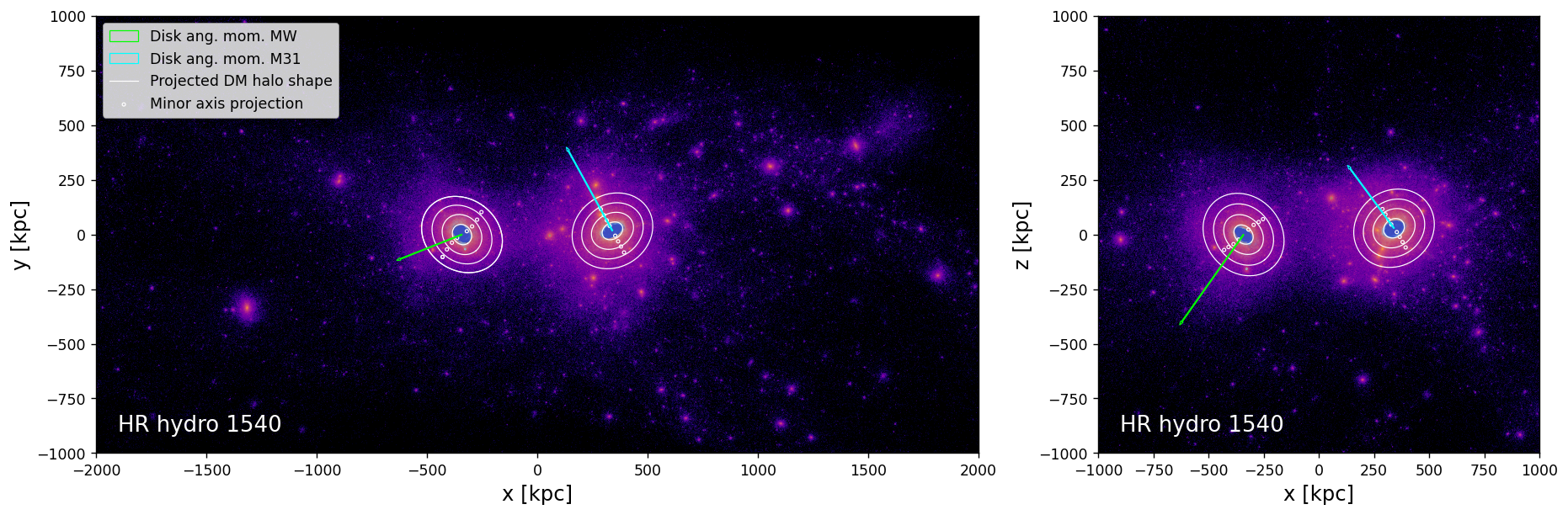}
	\end{subfigure}\vspace{-0.4em}
	\caption{\small \textbf{Top row:} \small Present-day snapshot of HR hydro simulation 720, showing the Milky Way (left) and M31 (right) in $(x, y)$ (4x2x2 Mpc volume) and $(x, z)$ (2x2x2 Mpc volume) projections. The white ellipsoids show the  {shapes projected onto the $(x,y)$ and $(x,z)$ plane} at $r$=50, 100, 150 and 200~kpc, open circles mark the projected minor axis at each radius. The projected direction of $\mathbf{L}_{\rm disk, MW}$ (lime arrow) and $\mathbf{L}_{\rm disk,M31}$ (cyan arrow) are overplotted. Dark matter particles are colorcoded by $\log_{10}$ density, star particles are colorcoded by age (young in blue, old in red). \textbf{Second - Fourth row}: Same as the top row, but showing HR hydro simulation 910, 1530, and 1540, respectively.}
	\label{fig:hydro_LSS_hr}
\end{figure}

\begin{figure}
	\centering
	\begin{subfigure}{0.82\textwidth}
		\centering
		\includegraphics[width=\linewidth]{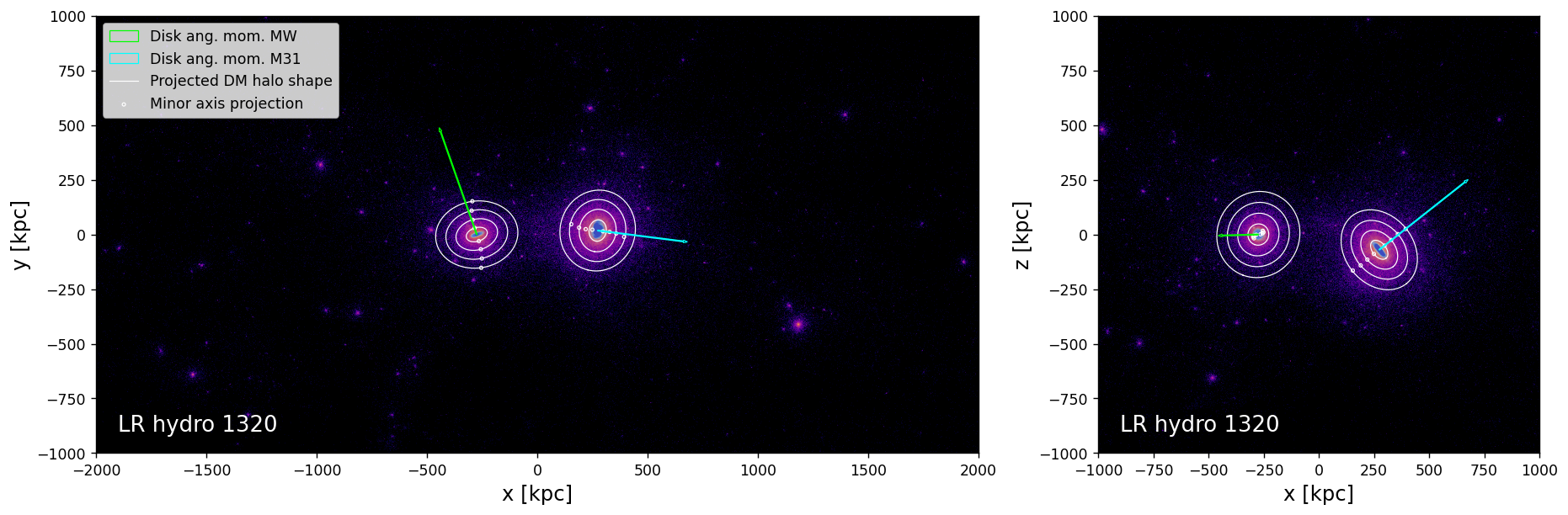} 
	\end{subfigure}\vspace{-0.4em}
	\begin{subfigure}{0.82\textwidth}
		\centering
		\includegraphics[width=\linewidth]{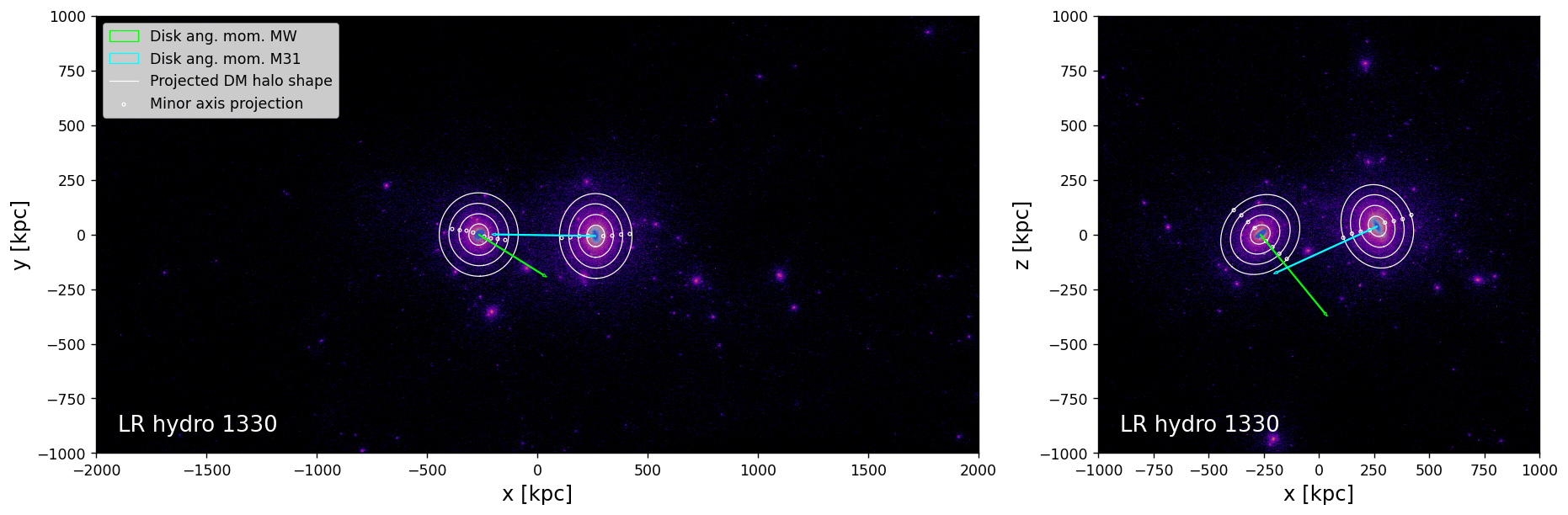}
	\end{subfigure}\vspace{-0.4em}
	\begin{subfigure}{0.82\textwidth}
		\centering
		\includegraphics[width=\linewidth]{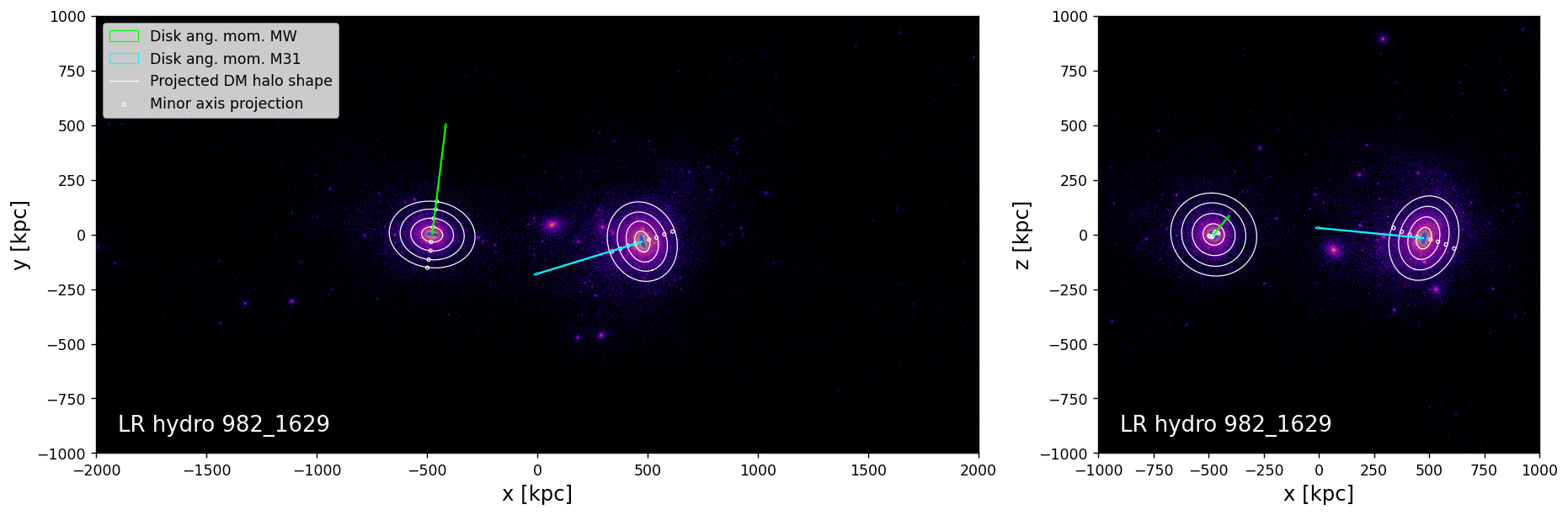}
	\end{subfigure}\vspace{-0.4em}
    \begin{subfigure}{0.82\textwidth}
		\centering
		\includegraphics[width=\linewidth]{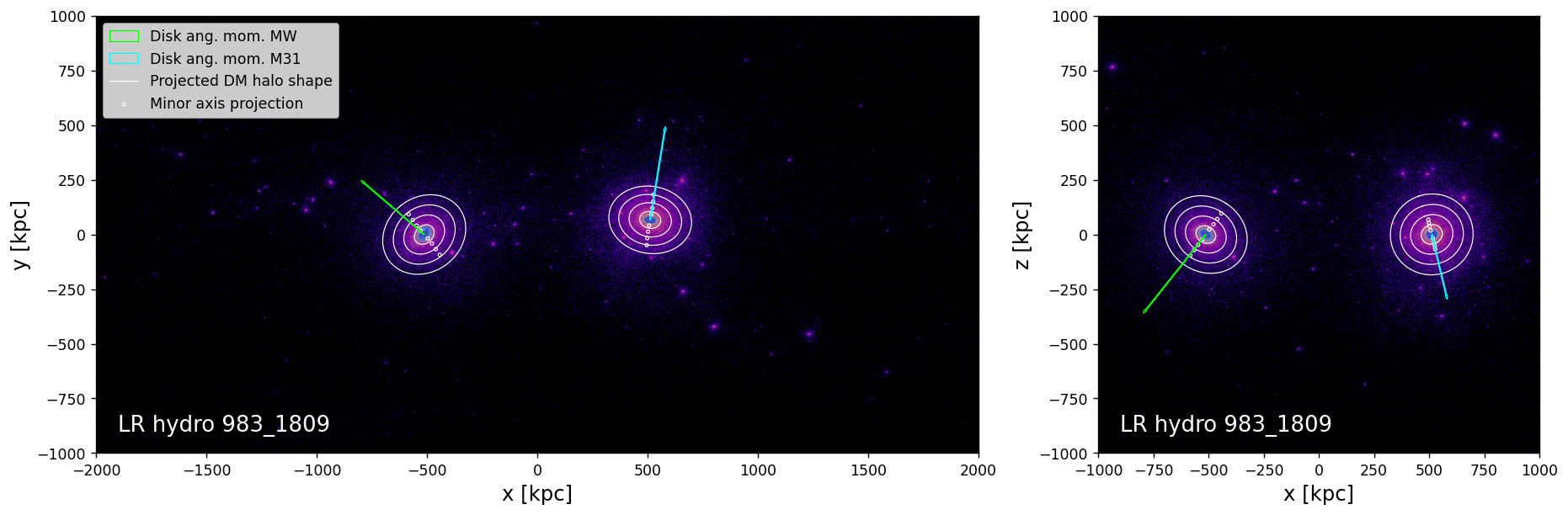}
	\end{subfigure}\vspace{-0.4em}
     \begin{subfigure}{0.82\textwidth}
		\centering
		\includegraphics[width=\linewidth]{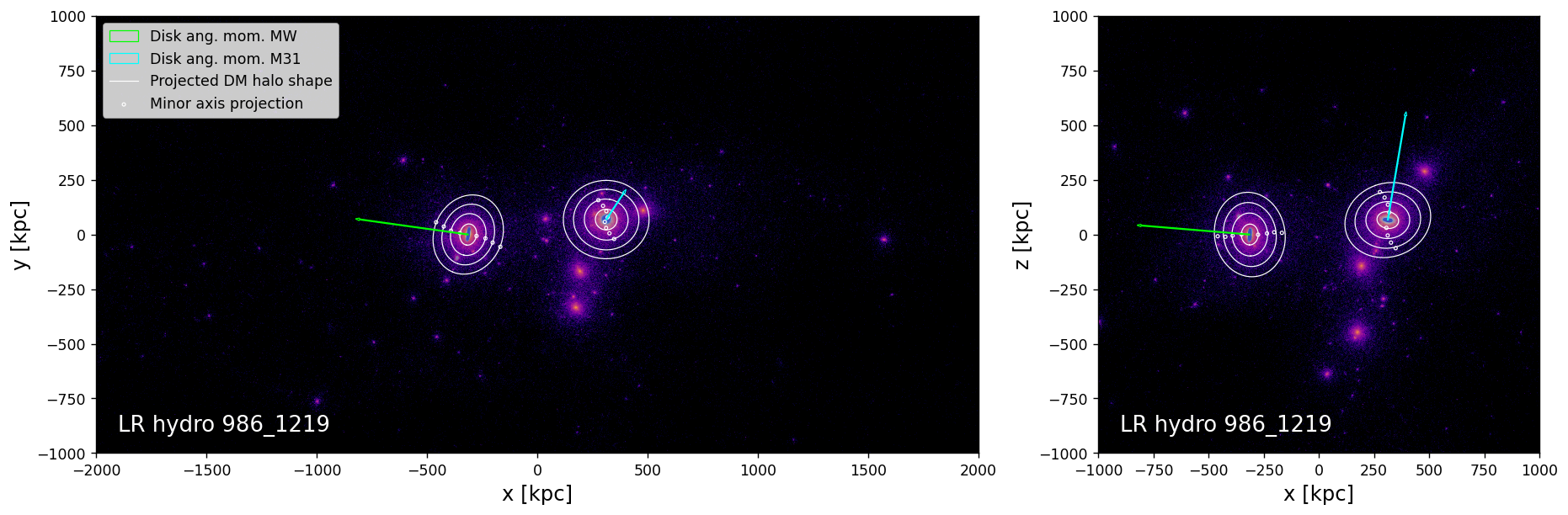}
	\end{subfigure}\vspace{-0.4em}
	\caption{\small Same as Fig. \ref{fig:hydro_LSS_hr}, but showing the LR hydro simulations 1320, 1330, and 982\_1629, 983\_1809, and 986\_1219 respectively.}
	\label{fig:hydro_LSS_lr}
\end{figure}

For completeness, we show the projected shapes and $\mathbf{L}_{\rm disk}$ directions for the four HR hydrodynamical runs in Fig. \ref{fig:hydro_LSS_hr} and for the five LR hydrodynamical runs in Fig. \ref{fig:hydro_LSS_lr}. 

\end{appendix}

\end{document}